\documentclass{article}

\usepackage{arxiv}

\usepackage[utf8]{inputenc} % allow utf-8 input
\usepackage{enumitem}  % item separation 
\usepackage[T1]{fontenc}    % use 8-bit T1 fonts
\usepackage{fancyhdr}
\usepackage{mathtools}
\usepackage{float}
\usepackage{amsmath}
\usepackage{hyperref}       % hyperlinks
\usepackage{url}            % simple URL typesetting
\usepackage{booktabs}       % professional-quality tables
\usepackage{amsfonts}       % blackboard math symbols
\usepackage{nicefrac}       % compact symbols for 1/2, etc.
\usepackage{microtype}      % microtypography
\usepackage{graphicx}
\graphicspath{{images/}}
\usepackage[square,sort,comma,numbers]{natbib}
\usepackage{doi}

\title{Aerospace Sliding Mode Control Toolbox: Relative Degree Approach with Resource Prospector Lander and Launch Vehicle Case Studies}

\author{ \href{https://orcid.org/0000-0001-5496-7901}{\includegraphics[scale=0.06]{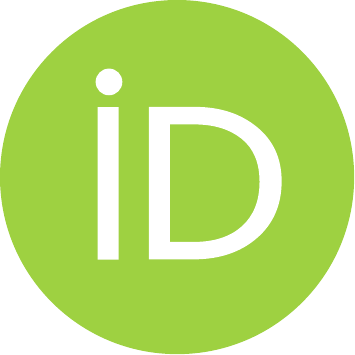}\hspace{1mm}Sai ~Kode} \\
	Department of Electrical and Computer Engineering\\
	The University of Alabama in Huntsville\\
	Huntsville, AL 35899 \\
	\texttt{sk0041@uah.edu} \\
	
	\And
	\href{https://orcid.org/0000-0000-0000-0000}{\includegraphics[scale=0.06]{orcid.pdf}\hspace{1mm}Yuri B.~Shtessel} \\
	Department of Electrical and Computer Engineering\\
	The University of Alabama in Huntsville\\
	Huntsville, AL 35899 \\
	\texttt{shtessy@uah.edu} \\
	
	\And
	\href{https://orcid.org/0000-0000-0000-0000}{\includegraphics[scale=0.06]{orcid.pdf}\hspace{1mm}Arie ~Levant} \\
    Department of Applied Mathematics\\
	Tel-Aviv University\\ 
	Tel Aviv-Yafo, Israel\\
	\texttt{levant@post.tau.ac.il} \\
	
	\And
	\href{https://orcid.org/0000-0000-0000-0000}{\includegraphics[scale=0.06]{orcid.pdf}\hspace{1mm}John ~Rakoczy} \\
	Flight Controls Branch\\
	NASA/Marshall Space Flight Center\\
	Huntsville, AL 35801\\
	\texttt{john.m.rakoczy@nasa.gov} \\
	
	\And
	\href{https://orcid.org/0000-0000-0000-0000}{\includegraphics[scale=0.06]{orcid.pdf}\hspace{1mm}Mike ~Hannan} \\
	Flight Controls Branch\\
	NASA/Marshall Space Flight Center\\
	Huntsville, AL 35801\\
	\texttt{mike.r.hannan@nasa.gov} \\
	
	\And
	\href{https://orcid.org/0000-0000-0000-0000}{\includegraphics[scale=0.06]{orcid.pdf}\hspace{1mm}Jeb ~Orr} \\
	Flight Systems Group\\
	Mclaurin Aerospace\\
    Huntsville, AL 35804\\
	\texttt{jeb.orr@mclaurin.aero} \\

}

\date{}
\begin{document}
\maketitle

\begin{abstract}
	Sliding mode control and observation techniques are widely used in aerospace applications, including aircraft, UAVs, launch vehicles, missile interceptors, and hypersonic missiles. This work is dedicated in creating a MATLAB based sliding mode controller design and simulation software toolbox that aims to support aerospace vehicle applications. An architecture of the aerospace sliding mode control toolbox (\emph{SMC Aero}) using the relative degree approach is proposed. The \emph{SMC Aero} libraries include first order sliding mode control (1-SMC), second order sliding mode control (2-SMC), higher order sliding mode (HOSM) control (either fixed gain or adaptive), as well as higher order sliding mode differentiators. The efficacy of the \emph{SMC Aero} toolbox is confirmed in two case studies: controlling and simulating resource prospector lander (RPL) soft landing on the Moon and launch vehicle (LV) attitude control in an ascent mode.
\end{abstract}


\section{Introduction}
Conventional Sliding Mode Control (SMC or 1-SMC) \cite{ref1,ref2,ref3} and Higher Order Sliding Mode Control (HOSMC) \cite{ref1,ref4,ref5} are powerful control techniques for achieving a finite time convergence in nonlinear systems in the presence of bounded perturbations. The relative degree approach \cite{ref1,ref4,ref5} allows designing SMC/HOSMC without knowledge of the system's mathematical model, while reducing the smooth unmatched disturbances to matched ones. Also, the relative degree approach allows achieving a continuous/smooth control by artificial increase of relative degree without compromising the system's robustness to smooth bounded perturbations. The numerous relative degree based adaptive SMC/HOSMC algorithms (see, for instance, \cite{ref6,ref7,ref21,ref22,ref23,ref24,ref25,ref26} allows self-tuning controller design in the presence of bounded disturbances with unknown bounds. The HOSM differentiators/observers (HOSMO) \cite{ref1,ref4,ref37} facilitate the HOSMC design.

SMC/HOSMC aerospace applications include missile interceptor's guidance and control against maneuvering targets \cite{ref11,ref16,ref17}, aircraft, UAVs, and launch vehicle's control in the presence of model uncertainties and actuator malfunctions \cite{ref9,ref10,ref12,ref13,ref14,ref18,ref19,ref41,ref42}, as well as hypersonic vehicle's control \cite{ref15,ref22}. The numerous applications of SMC/HOSMC to aerospace vehicle control calls for creating a toolbox to facilitate a design and simulation of robust finite time convergent aerospace control systems in the presence of bounded perturbations. The reported in this work relative degree approach-based \emph{SMC Aero} toolbox features a variety of SMC/HOSMC/Observation algorithms, including the identification on Practical Relative Degree (PRD). Note that the use of the exact Differentiation/Observation algorithms could replace expensive sensors in feedback control. The PRD identification can also be carried out easily by using the \emph{SMC Aero} toolbox, which in the future may extend its applications to the field of medicine, robotics and machine-learning environments.

The SMC/HOSMC primary features are implemented in a single platform, the MATLAB environment. This approach makes the toolbox an effective engineering resource for addressing multiple aerospace control problems. Collections of various built-in functions, a live scripting programmable environment, masking features (that allow encapsulating the block logic by hiding the block data beneath a simpler interface), and the ease of development using a graphical block model justifies the selection of MATLAB/Simulink over other engineering platforms in designing the toolbox. The toolbox architecture is marked by its ability to design the SMC/HOSMC based on performance specifications. These performance specifications are embedded in the function library and create an edge over other control toolboxes that primarily address only analysis, but not design, capabilities. The toolbox provides the user with an algorithm map, e.g., a well-organized cluster of potential design and analysis methods. Each block includes a detailed description embedded in the mask so that documentation is accessible directly within the Simulink library. Implementation of wide range of algorithms of different relative degrees enables an analyst to assess multiple design opportunities within a unified analysis platform. The process of designing/analyzing the SMC/HOSMC systems accommodates system models expressed either in state space or transfer function representations.

The main contribution of the paper is in presenting a novel \emph{SMC Aero} toolbox that includes a variety of SMC/HOSMC/HOSMO structures and algorithms for aerospace applications. The efficacy of the \emph{SMC Aero} toolbox is verified using two different case studies. One of the case studies implements control design for the soft lunar landing of a Resource Prospector Lander (RPL). The second demonstrates launch vehicle (LV) attitude control design and simulation. Two specific libraries, respectively, facilitate the SMC/HOSMC implementation of the pulse-off control strategy for RPL \cite{ref8} and adaptive continuous SMC for LV attitude control \cite{ref30}. One of the strategies implemented in \emph{SMC Aero} toolbox includes providing RPL soft landing on the Moon using the pulse-off control only with a certain switching frequency and duration of pulses in the presence of center of mass misalignment and actuator/thruster malfunctions. The other control strategy tested by \emph{SMC Aero} toolbox includes robust control of the LV attitude in the presence of flexible modes, actuator dynamics and bounded perturbations.

\section{Motivating Examples}

\subsection{Resource Prospector Lander Control}

\begin{figure}
    \centering
	%\fbox{\rule[-.5cm]{4cm}{4cm} \rule[-.5cm]{4cm}{0cm}}
	\includegraphics[scale=1]{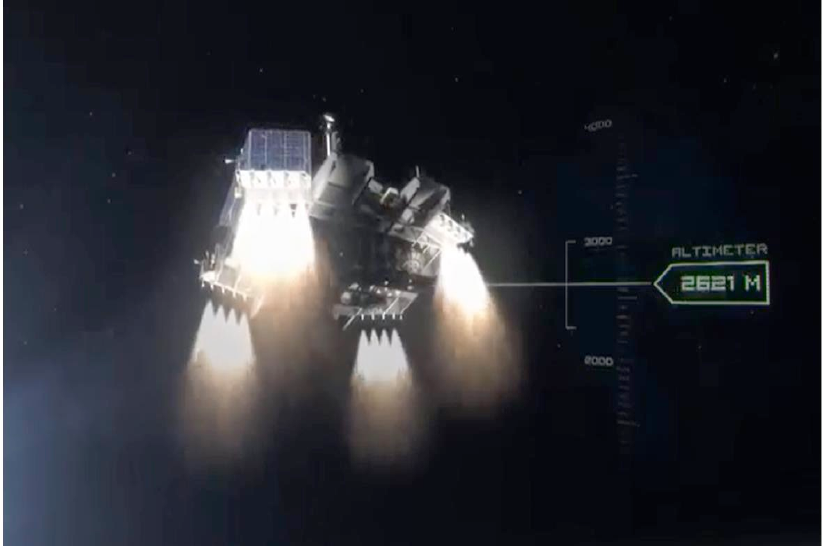}
	\caption{RP Lander Active Descent Control thrusters \cite{ref8}}
	\label{fig:fig1}
\end{figure}

A pulse-off attitude and position/velocity controller that allow pinpoint soft landing on the moon of the Resource Prospector Lunar Lander (RPL) \cite{ref8} (Figure~\ref{fig:fig1}) in the presence of model perturbations using only thrusters is considered.

The control functions are to be designed in terms of pitch and yaw control torques, as well as an axial force with consequential control allocation of twelve descent and twelve attitude control thrusters. An Adaptive Conventional Sliding Mode Control (1-SMC) algorithm is a good candidate for the attitude and descent pulse-off controller design to achieve RPL soft landing in the presence of center of mass misalignment and actuator/thruster malfunctions.

\subsection{Launch Vehicle Control}
\begin{figure}[H]
    \centering
	\includegraphics[scale=1]{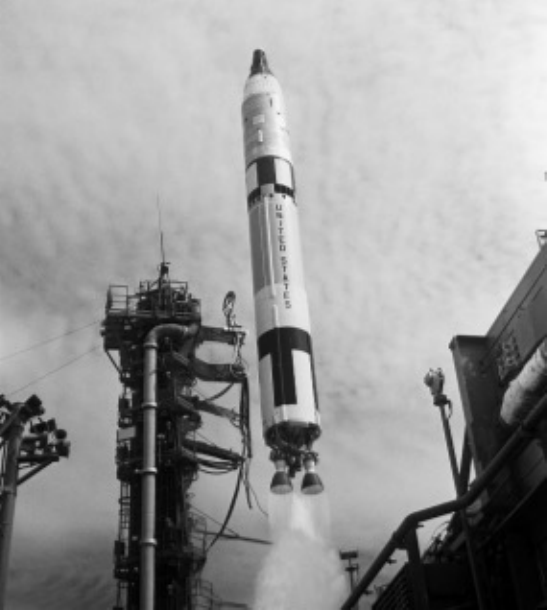}
	\caption{Gemini-Titan launch vehicle: symmetric actuation \cite{ref40}}
	\label{fig:fig2}
\end{figure}

Attitude control of a launch vehicle (LV) (Figure~\ref{fig:fig2}) \cite{ref9,ref32}, whose dynamics consists of a second order aerodynamically unstable air frame with a third-order servo actuator in the presence of structural flexibility and bounded perturbations is considered. The control functions are to be designed in terms of gimbal angle deflections under quasi-static thrust. The Adaptive Continuous Higher Order Sliding Mode Control (AHOSMC) algorithm is a good candidate for robust, accurate attitude controller design in the presence of bounded perturbations. Note that practical relative degree identification is a very important issue to be resolved prior to designing the AHOSMC controller.

The RPL pulse-off 1-SMC controller design, the LV robust accurate AHOSMC controller design, and the control system simulations are facilitated by a MATLAB-based \emph{SMC Aero} toolbox proposed and developed in this paper.

\section{Problem Statement}

\subsection{Requirements for the SMC Aero Architecture}

In this work, the control architecture is based on the relative degree approach used in the aerospace sliding mode control toolbox (\emph{SMC Aero}) that facilitates

\begin{itemize}
\item
  accurate and precise controller design for aerospace vehicles in the presence of the bounded perturbations,
\item
  simulations of the sliding mode control systems of aerospace vehicle control systems
\end{itemize}

is to be developed and verified on case studies, specifically, RPL and LV control and simulations.

The \emph{SMC Aero} toolbox architecture includes

\begin{itemize}
\item
  a variety of high frequency switching, continuous, fixed-and adaptive-gain sliding mode and higher order mode control algorithms to facilitate the aerospace vehicle control design and simulations, including
\item
  conventional, second and higher order sliding mode control and differentiation/observation algorithms (1-SMC, 2-SMC, and n-SMC or HOSMC),
\item
  the practical relative degree identification algorithm,
\item
  numerous service blocks and subroutines.
\end{itemize}

The contributions of the paper are as follows:

\begin{itemize}
\item
  The architecture of the toolbox based on the relative degree approach MATLAB-based aerospace sliding mode control toolbox (\emph{SMC Aero}) that facilitates the aerospace vehicle controller designs and simulations are proposed and developed for the first time,
\item
  The developed \emph{SMC Aero} toolbox is verified on two cases study:
  controlling RPL in the powered descent mode on the Moon and controlling attitude of the LV in the ascent mode, both in the presence of the bounded perturbations.
\end{itemize}

\subsection{Mathematical models of aerospace vehicles}

Dynamics of an aerospace vehicle are described by a generic system of nonlinear differential equations

\begin{equation}\label{eq:1}
    \begin{cases}
      \dot{x} = f(x,t)+B(x,t)u+D(x,t)\varphi(t)\\
       y = h(x,t)
    \end{cases}       
\end{equation}

where \(x \in R^{n}, y \in R^{n}, u \in R^{m}, m \leq n, \varphi \in R^{k} \) are state, control, output, and perturbation vectors respectively; the smooth enough matrices \(B(x,t), D(x,t)\) are of corresponding dimensions and can be partially known. The perturbation vector \(\varphi\) (and sometimes its consecutive derivatives) is norm bounded.

An output tracking problem \(||e|| \to 0\) (where \(e = y_{c}- y\), and \(y_{c}\) is a smooth enough output command generated online) is to be addressed by 1-SMC/2-SMC/HOSMC control \(u\) \cite{ref1} either in finite time, asymptotically, or in terms of uniform ultimate boundedness \cite{ref31}.

\begin{equation}\label{eq:2}
     \begin{bmatrix}
           y_{1}^{(r_{1})} \\
           y_{2}^{(r_{2})} \\
           \vdots \\
           y_{m}^{(r_{m})}
         \end{bmatrix}=
         \begin{bmatrix}
               \bar{f_{1}}(x,t) \\
               \bar{f_{2}}(x,t) \\
               \vdots \\
               \bar{f_{m}}(x,t) 
         \end{bmatrix}+
         \bar{D}(x,t)
         \begin{bmatrix}
               \bar{\psi}_{1}(t) \\
               \bar{\psi}_{2}(t) \\
               \vdots \\
               \bar{\psi}_{m}(t) 
         \end{bmatrix}+
         G(x,t)
         \begin{bmatrix}
               u_{1} \\
               u_{2} \\
               \vdots \\
               u_{m} 
         \end{bmatrix}
\end{equation}

where 

\begin{equation}\label{eq:3}
G(x,t) = 
\begin{bmatrix}
g_{11}(x,t) & g_{12}(x,t) & \cdots & g_{1m}(x,t) \\
g_{21}(x,t) & g_{22}(x,t) & \cdots & g_{2m}(x,t) \\
\vdots  & \vdots  & \ddots & \vdots  \\
g_{m1}(x,t) & g_{m2}(x,t) & \cdots & g_{mm}(x,t) 
\end{bmatrix}, \: det[G(x,t)] \neq 0
\end {equation}

\begin{equation*}
\bar{D}(x,t)=diag\{d_{ii}(x,t)\},i=1,2,....,m;\: r_{t}=\sum_{i=1}^{m}r_{i},\: r_{t} \leq n
\end{equation*}

Assume \\
(A1) the norm bounded matrix \(G(x,t)\) can be presented as \(G = G_{0}+\Delta G, \: det(G_{0})\neq 0\), where the known matrix \(G_{0}\) diagonally dominates the unknown matrix \(\Delta G\), in the sense that

\begin{gather*}
 G = G_{0}(I+G_{0}^{-1}\Delta G), \tilde{G}=G_{0}^{-1}\Delta G = (\tilde{g}_{ij}),\\
\exists \: \gamma \in (0,1):\forall i \sum_{j} |{\tilde{g}_{ij}}|\leq \gamma
\end{gather*}

The input-output tracking error dynamics are derived as

\begin{equation}\label{eq:4}
    \begin{bmatrix}
    e_{1}^{(r_{1})} \\[0.1cm]
    e_{2}^{(r_{2})} \\[0.1cm]
    \vdots \\
    e_{m}^{(r_{m})} \\
    \end{bmatrix} = 
    \begin{bmatrix}
    \xi_{1}(x,t) \\
    \xi_{2}(x,t) \\
    \vdots \\
    \xi_{m}(x,t) \\
    \end{bmatrix} -
    G(x,t)
    \begin{bmatrix}
    u_{1} \\
    u_{2} \\
    \vdots \\
    u_{m}
    \end{bmatrix}
\end{equation}       

where 

\begin{equation}\label{eq:5}
     \begin{bmatrix}
    \xi_{1}(x,t) \\
    \xi_{2}(x,t) \\
    \vdots \\
    \xi_{m}(x,t) \\
    \end{bmatrix} = 
    \begin{bmatrix}
           y_{c}^{(r_{1})} \\
           y_{c}^{(r_{2})} \\
           \vdots \\
           y_{c}^{(r_{m})}
         \end{bmatrix} -
           \begin{bmatrix}
               \bar{f_{1}}(x,t) \\
               \bar{f_{2}}(x,t) \\
               \vdots \\
               \bar{f_{m}}(x,t) 
         \end{bmatrix}-
         \bar{D}(x,t)
         \begin{bmatrix}
               \bar{\psi}_{1}(t) \\
               \bar{\psi}_{2}(t) \\
               \vdots \\
               \bar{\psi}_{m}(t) 
         \end{bmatrix}
\end{equation}

Note that input-output dynamics format in (\ref{eq:2})-(\ref{eq:5}) allows considering the control system as a "black box" with a minimal knowledge about the perturbations.

\section{SMC Aero Control Toolbox}

\subsection{Architecture of SMC Aero toolbox}

\begin{figure}[H]
    \centering
    \includegraphics[width=8cm]{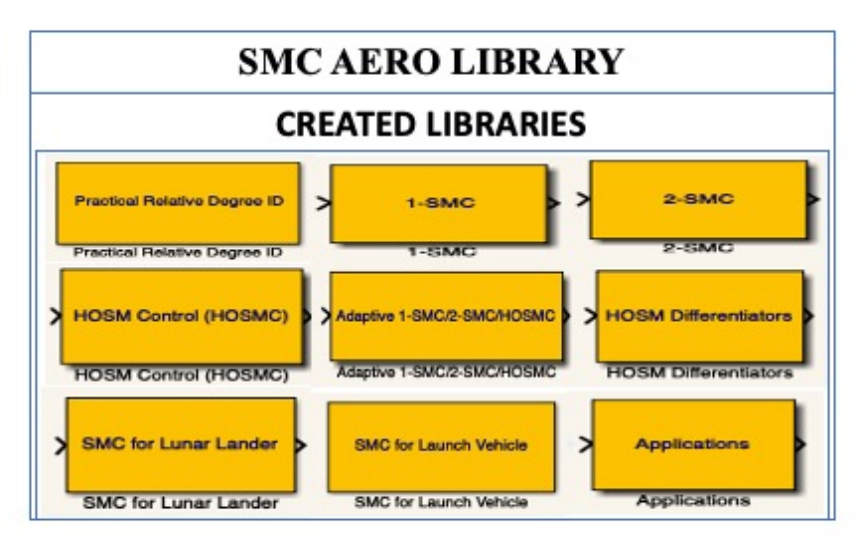}
    \caption{Structure of the SMC Aero Toolbox}
    \label{fig:fig3}
\end{figure}

A current first version of a proposed MATLAB-based architecture/structure of the \emph{SMC Aero} toolbox presented in (Figure~\ref{fig:fig3}) (a preliminary version is discussed in \cite{ref29} and presented for possible users in \cite{ref30}) contains the following libraries:

\begin{itemize}
\item
  Practical Relative Degree ID
\item
  1-SMC
\item
  2-SMC
\item
  HOSM Control (HOSMC)
\item
  Adaptive 1-SMC/2-SMC/HOSMC
\item
  HOSM Differentiators
\item
  SMC for Lunar Lander (RPL)
\item
  SMC for Launch Vehicle (LV)
\item
  Applications: SMC of Double Integrator, and DC Motor
\end{itemize}

It is worth reminding that that created the \emph{SMC Aero} toolbox supports the design and simulation of the aerospace vehicles controlled by 1-SMC/2-SMC/HOSMC controllers, including continuous and adaptive ones.

\subsection{Practical Relative Degree ID Library}

The definition of relative degree can be presented as \cite{ref31}

\textbf{Definition 1.} Consider a smooth SISO dynamic system (\ref{eq:1}) with \(x \in R^{n}, y \in R^{n}, u \in R^{1}\). If \(y^{(i)}\) independent of control \emph{u} for all \( i = 1,2,...,r-1\), where \(r > 0\) is an integer number, but \(\frac{\partial }{\partial u}y^{r}\)does not vanish in a domain \( (x,t) \in \Theta \subset \in R^{n+1}\), then \(r\) is called the \emph{relative degree}.

Note that using the concept of relative degree, all smooth enough unmatched perturbations become matched in the output dynamics (\ref{eq:5}) and therefore can be easily compensated by SMC/HOSM controllers. Practical relative degree ID algorithms were studied in \cite{ref27,ref28}. In the \emph{SMC Aero} toolbox, the algorithm presented in \cite{ref28} that is based on the idea of feeding a step function to the system's input, and then differentiating the output is coded. The following informal description is based on the rigorous definition presented in \cite{ref28}. If \(\text{u~}\epsilon\ \mathbb{R}^{1}\) is a step unit function fed to the input of SISO system (\ref{eq:1}), and the output \(\text{y~}\epsilon\ \mathbb{R}^{1}\) contains a step-like signal after \(r_{p}\) differentiations, then \(r_{p}\) is called Practical Relative Degree (PRD).

Note that \begin{itemize}
\vspace{-0.2cm}\item \(r_{p}\) can be both less or larger than \(r\),
\vspace{-0.2cm}\item The proposed algorithm can be implemented on simulations (given the system's mathematical model) or experimentally.
\end{itemize}

\subsection{Practical Relative Degree ID Implementation}

Informally, PRD, \(r_{p}\) is defined in SISO systems as a smallest order of the system's output derivative that contains the control input-like signal in a domain containing the equilibrium point. In \emph{SMC Aero} toolbox the PRD identification is accomplished as follows. It is assumed that PRD \(r_{p}\) of SISO dynamic system (\ref{eq:1}) exists but is unknown. In order to identify PRD \(r_{p}\) the following algorithm \cite{ref28} is used

\emph{Step1.} A Heaviside step function
\begin{equation}\label{eq:6}
    1(t-\tau)= 
    \begin{cases}
      1, \enspace if \: t\geq \tau \\
      0, \enspace otherwise.
    \end{cases} 
\end{equation}
is applied to the input of the SISO dynamic system.

\emph{Step 2.} The SISO system output \(y\) (sometimes this is a sliding variable \(\sigma\)) is differentiated multiple times using HOSM differentiator with a finite convergent time until discontinuity (or a steep change) is observed in the \({r_{p}}^{th}\) derivative or, consequently, a sharp change of a slope appears in the previous derivative \(y^{(r_{p}-1)}\) then \( PRD = r_{p}\). Algorithm handlers will stop the differentiation only when they observe a steep change in a corresponding derivative of the system's output, otherwise iteration will be carried forward without any stop. The implemented algorithms implement high-order differentiation robust to measurement noises.

\begin{figure}[H]
    \centering
    \includegraphics[width=8cm]{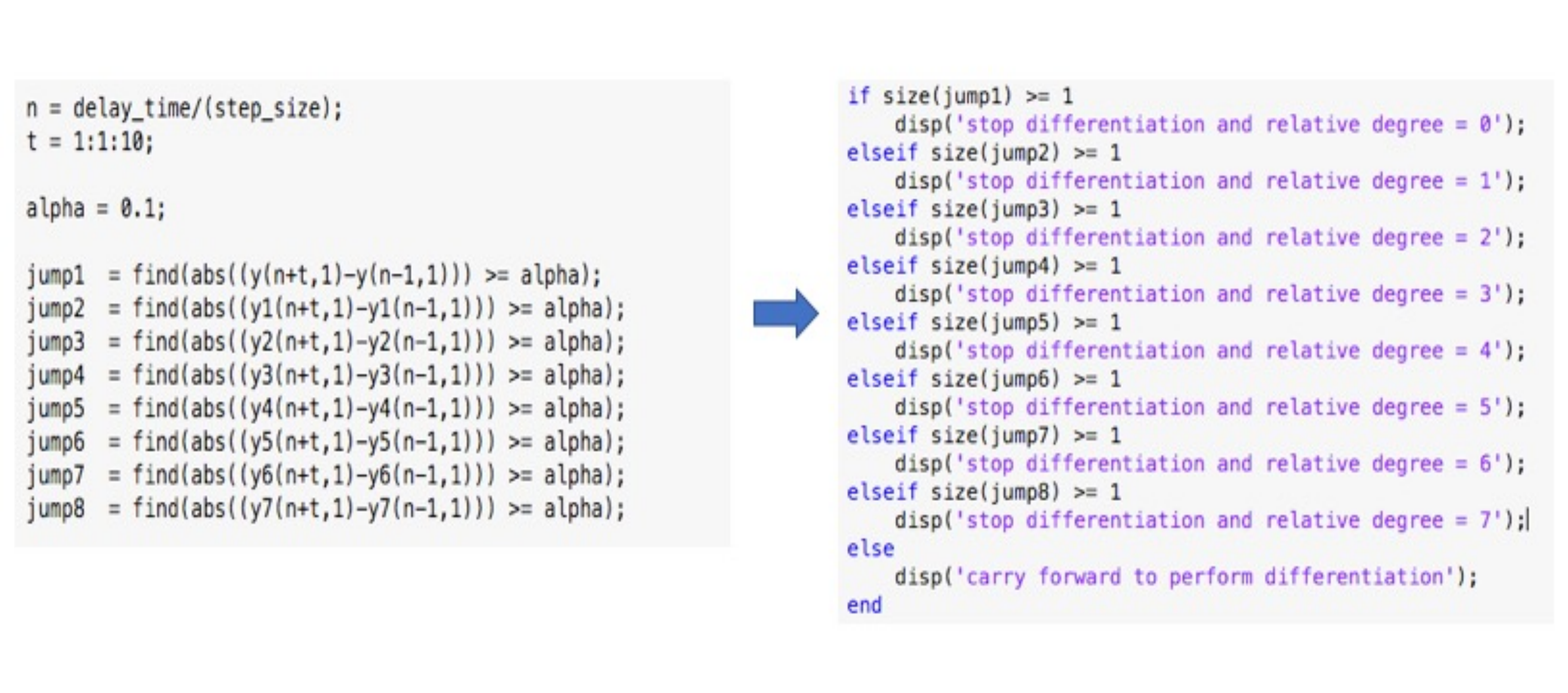}
    \caption{Live MATLAB editor code of a performance analyzer}
    \label{fig:fig4}
\end{figure}

\begin{figure}[H]
    \centering
    \includegraphics[scale=1]{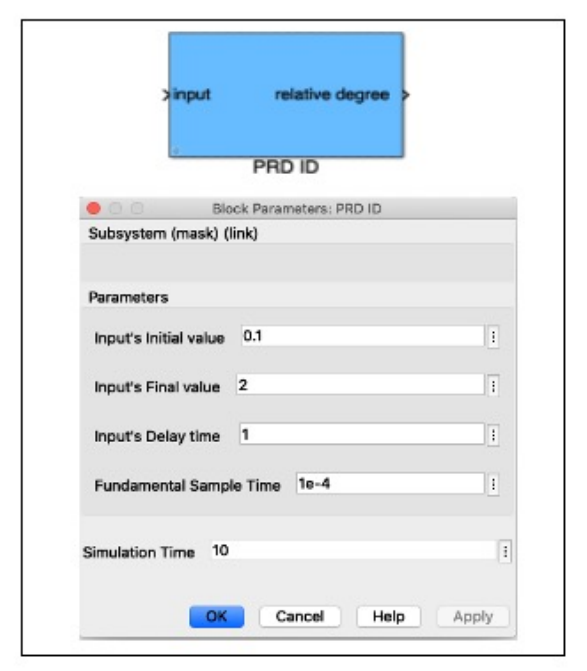}
    \caption{Practical Relative Degree ID block with an embedded mask feature}
    \label{fig:fig5}
\end{figure}

The PRD ID algorithm is implemented in a relative degree ID library of the \emph{SMC Aero} toolbox. Note that a crucial part of PRD ID code is a \emph{performance analyzer} that monitors the transient response and conveys the message to the algorithm handler, whether any steep change (or sudden change in a slope) of the transient response occurs. The performance analyzers are designed based on the analysis of a neighborhood of \(y(\tau)\), where \( t = \tau \) is a moment of the Heaviside input application. The MATLAB iterative code, which was embedded in the PRD ID algorithm, and PRD ID block with embedded mask feature are demonstrated in Figures~\ref{fig:fig4} and \ref{fig:fig5} respectively.

\subsection{Practical Relative Degree ID: algorithm verification}

The efficacy of the PRD ID algorithm implemented in the \emph{Practical Relative Degree ID} library, the \emph{SMC Aero} toolbox is verified on the launch vehicle in ascent mode, whose open loop (uncompensated) simplified input-output transfer function is defined as \cite{ref33}

\begin{equation}\label{eq:7}
\frac{\theta(s)}{\beta(s)} = \frac{a_{2}s^{2} + a_{1}s^{1} + a_{0}}{{b_{7}s^{7} + b_{6}s^{6} + b_{5}s^{5} + b}_{4}s^{4} + b_{3}s^{3} + b_{2}s^{2} + b_{1}s^{1} + b_{0}}
\end{equation}

where, \(a_{0}, a_{1}, a_{2}\) are \(- 16870\), \(- 13.55\), \(- 117.1\) respectively. Similarly, \(b_{0}, b_{1}\text{...}b_{7}\) are \(- 9000\), \(- 2557\), \(22230\), \(6364\),\(\ \ 679,\ \ 72.39,\ 3.425,\ \ 0.2\ \) respectively, and \( \theta, \beta \) are a pitch and a thrust deflection angles (\emph{rad}) respectively.

The difference between the degree of the denominator and degree of the numerator gives system's relative degree \( r = 7-2 = 5\) \cite{ref31}.

Next, PRD is identified using the \emph{Practical Relative Degree ID} library from \emph{SMC Aero} toolbox assuming that the transfer function in (\ref{eq:7}) is not known for computing relative degree but used only for generating the output response. The system output \( y(t) = \theta(t) \) is obtained "experimentally" (actually via simulations) and then differentiated multiple times until

\begin{enumerate}
\vspace{-0.2cm}\item
  sharp change of a slope of simulated profile emerges; then the number of this derivative plus one gives the PRD;
\vspace{-0.2cm}\item
  a steep change (quasi-discontinuity) in a simulated profile is detected; then the number of this derivative equal to PRD
\end{enumerate}

\emph{The PRD ID library set up for the verification test:}

\begin{figure}[H]
    \centering
    \includegraphics[scale=1.25]{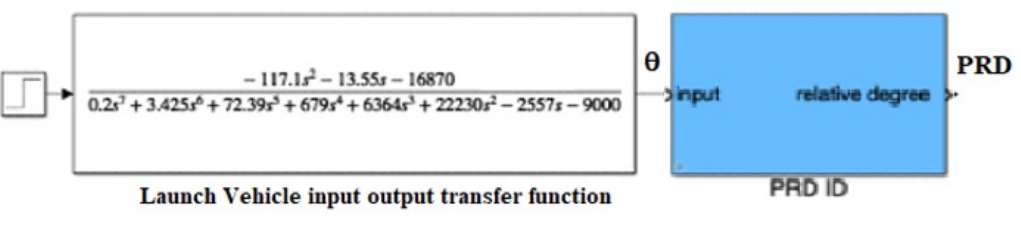}
    \caption{Simulation diagram for determining the PRD of the launch vehicle input-output dynamics using the \\ SMC Aero Toolbox}
    \label{fig:fig6}
\end{figure}

\begin{itemize}
\item
  Number of iterations \(n = 10^{4}\)
\item
  The parameter \(\alpha\) for the steep change identification \(\alpha = 0.1\) 
\item
  The initial time and time of the Heaviside function application are \( t_{0} = \tau = 1 s \) .
\item
  Method of integration: Euler with a fixed step size \( \Delta t = 10^{-4} s \)
\item
  Simulation diagram of PRD ID for the launch vehicle is presented in the (Figure~\ref{fig:fig6}).
\end{itemize}

The output \(y(t) = \theta (t)\) response and its derivatives are presented in the Figure~\ref{fig:fig6}, where the input was applied 

\begin{equation}\label{eq:8}
    \beta(t) = -17.5\cdot 1(t-1)+35.0\cdot 1(t-1.3)-35.0\cdot 1(t-1.85)
\end{equation}

\begin{figure}[H]
    \centering
    \includegraphics[scale=0.25]{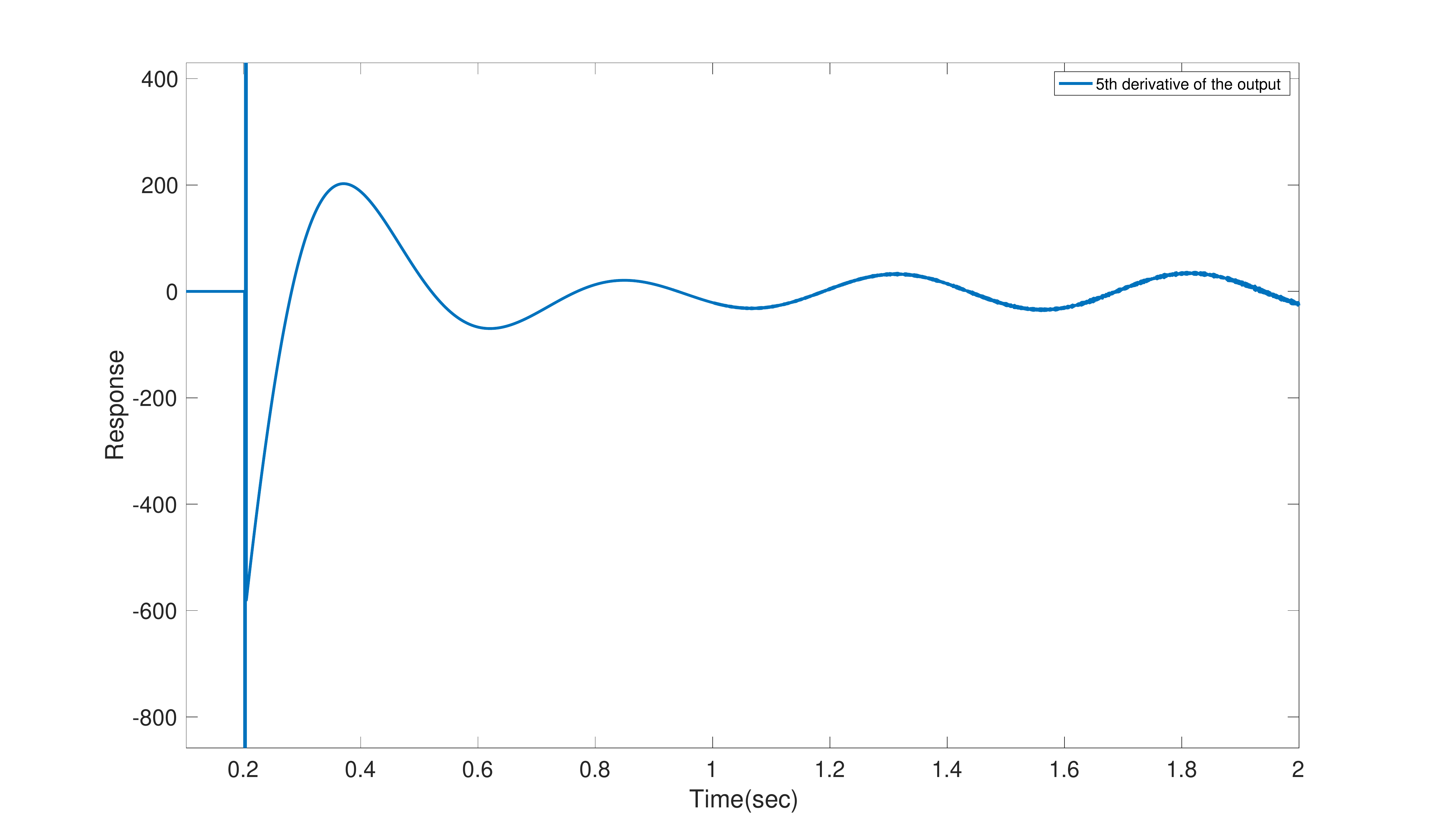}
    \caption{Step-like input vs. \(5^{th}\) derivative of the output}
    \label{fig:fig7}
\end{figure}

\emph{\textbf{Discussion}}: As soon as a steep change in the corresponding derivative of the system's output is detected by the performance analyzer, the PRD ID algorithms stops with printing a message. "The experimentally determined PRD is equal to 5". It is clear from (Figure~\ref{fig:fig7}) that the computed \(r_{p} = r = 5\) corresponds to the theoretical analysis of the transfer function (\ref{eq:7}). Note that PRD ID algorithm does not rely on the system dynamics equation (\ref{eq:7}) (that is used for verification purposes only) but is based exclusively on analysis of the system response.

\subsection{1-SMC Library}

This library consists of the blocks that address the output tracking problem with dynamics described by equations (\ref{eq:2}) and (\ref{eq:3}).

\subsubsection{MIMO 1-SMC Fixed Gain blocks}
MIMO 1-SMC fixed-gain blocks have been developed for \(m>1\). Assume that Assumption (A1) holds and vector relative degree \( \bar{r} = [r_{1}, r_{2},....,r_{m}]\) is given or identified.

Therefore, given vector relative degree \(\bar{r} = [r_{1}, r_{2},....,r_{m}]\) and the sliding variables are designed as

\begin{figure}[H]
    \centering
    \includegraphics[scale=1]{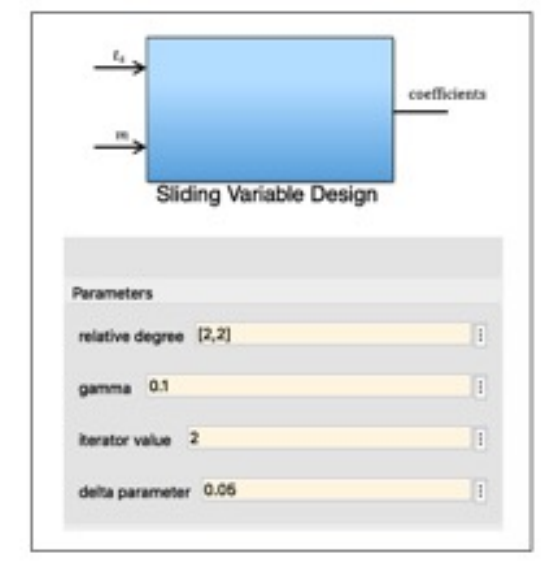}
    \caption{Sliding Variable Design block with a mask overview}
    \label{fig:fig8}
\end{figure}

\vspace{-0.3cm}\begin{gather} \nonumber
    \sigma = \left[\sigma_{1},\sigma_{2},....,\sigma_{m}\right]^{T} \\ \label{eq:9}
    \sigma_{i} = e_{i}^{(r_{i}-1)}+c_{i,r_{i}-2}e_{i}^{(r_{i}-2)}+...+c_{i,1}e_{i}^{(1)}+c_{i,0}e_{i}+c_{-i,1}\int e_{i}d\tau \\  \nonumber
    e_{i} = y_{ci}-y_{i}; \forall i = 1,2,...,m.
\end{gather}

where the coefficients are selected in accordance with ITAE criterion \cite{ref20} based on given settling time, and the integral term allows achieving an integral sliding mode \cite{ref1} and/or compensating the error bias that may occur upon discrete-time implementation. The sliding variable design block that computes the sliding variable coefficients in equation (\ref{eq:9}) is shown in Figure~\ref{fig:fig8}.

Note that the MIMO sliding variable dynamics can be derived as

\begin{equation}\label{eq:10}
    \dot{\sigma} = \bar{\xi}(x,t)-G_{0}(x)u
\end{equation}

where

\begin{gather}\label{eq:11}
    u = \left[u_{1},u_{2},....,u_{m}\right]^{T}, \: \bar{\xi} = \left[\bar{\xi}_{1},\bar{\xi}_{2},....,\bar{\xi}_{m}\right]^{T} \\ \nonumber
    \bar\xi_{i} = \bar\xi_{i}+e_{i}^{(r_{i})}+c_{i,r_{i}-2}e_{i}^{(r_{i}-1)}+....+c_{i,0}e_{i}^{(1)}+c_{-i,1}e_{i}
\end{gather}  

Assume

(A2) \( |\bar\xi_{i}(x,t)| \leq \bar{L}_{i}, ||\bar\xi_{i}(x,t)|| \leq \bar{L}, \: \bar{L}_{i}, \bar{L} > 0 \: \forall i = 1,2,...,m \) at least locally, i.e. \( \forall x \in \bar{\Omega}(x) \).

The discontinuous 1-SMC control \( u \in R^{m}\) that drives \( \sigma \to 0 \) is coded in two formats

(a) a unit vector format

\begin{equation}\label{eq:12}
    u = G_0^{-1}(x)v, \: v = \rho_{0}\frac{\sigma}{||\sigma||}, \: \rho_{0}>\bar{L}
\end{equation}

(b) a \( SIGN(\sigma)  = {[sign(\sigma_{1}), sign(\sigma_{2}),...,sign(\sigma_{m})]}^{T}\) format 

\begin{equation}\label{eq:13}
    u = G_0^{-1}(x)v, \: v = R\cdot SIGN(\sigma)
\end{equation}

with \( R = diag\{{\rho_{i}}\}, \rho_{i}>\bar{L}_{i} \: \forall i = 1,2,....,m \).

The continuous quasi-SMC is coded in a continuous sigmoid format, where

(a) a unit vector term \( \frac {\sigma}{||\sigma||} \) in (\ref{eq:12}) is replaced by \( \frac {\sigma}{||\sigma||+ \epsilon_{0}} \),

(b) \( sign(\sigma_{i})\) terms in (\ref{eq:13}) are replaced by \( \frac {\sigma_{i}}{|\sigma_{i}|+ \epsilon_{i}} \) with \( \epsilon_{0},\epsilon_{1}, ...., \epsilon_{m} > 0 \) being small real numbers.

Note that \cite{ref1}

\begin{itemize}
\item
  The fixed gains \( \rho_{i} > 0  \: \forall i = 0,1,...,m \) are required to be tuned, since the Lipschitz constants \( \bar{L},\bar{L}_{i} \: i = 1,2,..,m \) practically are not known. Another solution is based on the adaptive-gain SMC design and is considered later in this paper.
  
\item
  The continuous sigmoid control functions are \emph{quasi-SMCs,} since no convergence \(\sigma \rightarrow 0\ \) is guaranteed, but the convergence to a small domain around the origin is provided. This yields a partial loss of insensitivity to the perturbations of the system dynamics in the quasi-sliding mode.
\end{itemize}

\subsubsection{SISO 1-SMC Fixed Gain blocks} 
represent a particular, but a very important case with \(m = 1\). These blocks were coded separately from \emph{MIMO 1- SMC Fixed Gain blocks} and makes a subset of blocks of the SMC library.

\subsubsection{Adaptive 1-SMC blocks}
Assume that

(A3) in the assumption (A2) \( \bar{L},\bar{L}_{i} \: i = 1,2,..,m \) exist but are unknown.

In this case, when the perturbations are bounded with unknown bounds, the fixed gain 1-SMC may exhibit gain overestimation, which can lead to increased chattering. As a result, gain adaptation in 1-SMC may be needed if the perturbation bounds exist but are unknown.

The \emph{Adaptive 1-SMC blocks} include implementations that address the gain adaptation of SMC in equations (\ref{eq:12}) and (\ref{eq:13}).

\emph{MIMO 1-SMC Adaptive Gain blocks} were \emph{implemented and coded in the SMC Aero toolbox} for \(m > 1\). Two adaptive algorithms are implemented:

(a) \emph{The simple gain adaptation} laws \cite{ref6} were employed for an adaptive unit-vector control in (\ref{eq:12})

\begin{equation}\label{eq:14}
    \dot{\rho}_{0}=
     \begin{cases}
      \gamma_{0}||\sigma||, \enspace if \: ||\sigma||>\delta_{0} \\
      0, \qquad \enspace otherwise.
    \end{cases}
\end{equation}

and for adaptive vector \( SIGN(\sigma) \) based control (\ref{eq:13})

\begin{equation}\label{eq:15}
    \dot{\rho}_{i}=
     \begin{cases}
      \gamma_{i}|\sigma_{i}|, \enspace if \: |\sigma_{i}|>\delta_{i} \\
      0, \qquad \: otherwise.
    \end{cases}
\end{equation}

where, \( \gamma_{i}, \delta_{i} > 0 \: \forall i = 0,1,2...,m \) are the adaptation rate and the adaptation accuracy respectively.
Note that the control gains \( \rho_{i}\) are increased until a real sliding mode \(|| \sigma || \leq \delta_{0}\) or \(|\sigma_{i}| \leq \delta_{i}  \: \forall i = 0,1,2...,m \) get established, then \(\rho_{i} = const\).

Note that

\begin{itemize}
\item
  the gain adaptation laws in equations (\ref{eq:14}) and (\ref{eq:15}) are also applied to \emph{quasi-SMC} with the sigmoid approximation within the \emph{SMC Aero} toolbox;
\item
  the disadvantage of \emph{MIMO SMC Adaptive Gain blocks} in equations (\ref{eq:14}), (\ref{eq:15}) is in the fact that the adaptive gains cannot be reduced, and,
  therefore, can be overestimated.
\end{itemize}

Therefore, the advanced double layer gain adaption law \cite{ref26} that allows the gain reduction (non-overestimation) was also coded and implemented in the \emph{SMC Aero} toolbox.

(b) The \emph{advanced double layer gains adaptation law} for unit vector control in equation \ref{eq:12} was employed

\begin{equation}\label{eq:16}
    v = (k(t)+\eta)\frac{\sigma(t)}{||\sigma(t)||}
\end{equation}

with

\begin{equation}\label{eq:17}
    \delta(t) = k(t)-\frac{1}{\alpha}||\bar{v}_{eq}||-\epsilon
\end{equation}

where \( \bar{v}_{eq}\) can be obtained in real time by low pass filtering of the switching signal \(v\) while \(0 < \alpha < 1\) and \(\epsilon > 0\) are design scalars.

In the \emph{first layer} of the control gain \(k(t)\) adaptation the following adaptation law is proposed

\begin{equation}\label{eq:18}
    \dot{k}(t) = -\rho(t)sign(\delta(t))
\end{equation}

where, \(\rho(t) > 0\) is scalar adapted in the \emph{second layer} of the proposed adaptation algorithm is defined by

\begin{equation}\label{eq:19}
    \rho(t) = r_{0}+r(t), r_{0}>0
\end{equation}

with \(r_{0}\) to be a fixed positive (small) constant.

In the \emph{second layer} of the control gain \(k(t)\) adaptation the following adaptive law is proposed and coded for the gain \(\rho\left( t \right)\ \) in equation (\ref{eq:19}) as:

\begin{equation}\label{eq:20}
    \dot{r}(t) = 
      \begin{cases}
      \gamma|\delta(t)|, \enspace if \: |\delta(t)|>\delta_{0} \\
      0, \qquad \enspace \: otherwise.
    \end{cases}
\end{equation}

where, \(\delta_{0} > 0\) is a custom selected adaptation accuracy parameter.

Note that the advantage of \emph{MIMO Double Layer Adaptive Gain block in} equations (\ref{eq:16}) - (\ref{eq:20}) is in the fact that the adaptive control gains are
unlikely to be overestimated.

\begin{figure}[H]
    \centering
    \includegraphics[scale=1]{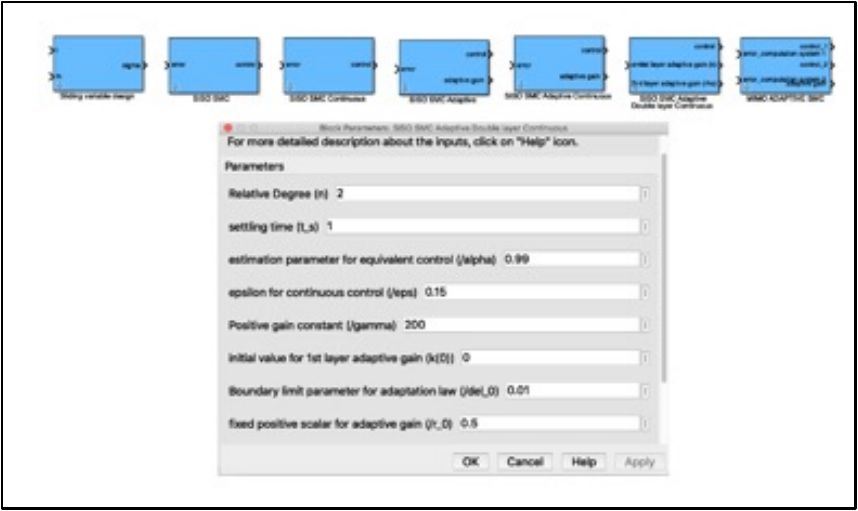}
    \caption{1-SMC library blocks with the continuous SISO SMC double layer mask}
    \label{fig:fig9}
\end{figure}

The 1-SMC library blocks with a mask for the Continuous SISO SMC double layer block are shown in Figure~\ref{fig:fig9}.

\subsection{2-SMC Library}
This library consists of the blocks that address the output tracking problem on 2-SMC of relative degree \(r = 1\) in nonlinear perturbed SISO systems
in equations (\ref{eq:2}) - (\ref{eq:5}), (\ref{eq:10}) with \(m = 1\).

The 2-SMC control \( u \in R^{1}\) that drives \(\sigma,\dot{\sigma} \rightarrow 0\) in finite time and keeps \(\sigma,\dot{\sigma} = 0\) for all consecutive times is coded in fixed-gain and adaptive-gain 2-SMC formats.

Assume that

(A4) There exists, at least locally, \emph{known} \( \tilde{L}_{1} > 0 \) so that \( |\bar{\dot{\xi}}(x,t)|\leq \tilde{L}_{1}\).

\subsubsection{Fixed gain 2-SMC blocks}

(a) \emph{Super Twisting (STW) controller block} \cite{ref1,ref34}

\begin{gather} \label{eq:21}
    u = G_{0}^{-1}(x,t)v \\ \nonumber
    v = \lambda|\sigma|^{1/2}sign(\sigma)+w, \dot{w}=\beta sign(\sigma) \\ \nonumber
\end{gather}

\vspace{-0.5cm} where \(\lambda\), \(\beta\) are fixed control gains \( \lambda = 1.5 \sqrt{\bar{L}_{1}}, \beta = 1.1 \bar{L}_{1}\). Note that the STW controller in (\ref{eq:21}) is a continuous controller, and \( G_{0}\) is a scalar function.

(b) \emph{Twisting (TW) controller block} \cite{ref1,ref34}

The TW controller is valid for the sliding variable dynamics of relative degree 2. That's why the TW control is designed in terms of control derivatives:

\vspace{-0.3cm}\begin{gather} \nonumber
    u = G_{0}^{-1}(x,t)\int \dot{v}dt \\ \label{eq:22}
    \dot{v} = -v, |v|\geq v_{m} \\ \nonumber
    \dot{v} = \alpha_{1}sign(\sigma)+\alpha_{2}sign(\dot{\sigma}), |v|<v_{m} 
\end{gather}

\vspace{-0.1cm}where \(v_{M}\) is an assigned control magnitude, \( \alpha_{1}, \alpha_{2} \) are fixed control gains that are to satisfy the conditions

\vspace{-0.3cm} \begin{equation}\label{eq:23}
    \alpha_{1}+\alpha_{2}-\tilde{L}_{1} > \alpha_{1}-\alpha_{2}+\tilde{L}_{1},\enspace \alpha_{1}-\alpha_{2}>\tilde{L}_{1}
\end{equation}

Note that

(a) both STW and TW controllers in (\ref{eq:21}), (\ref{eq:22}) are continuous controllers;

(b) the coefficients \( \lambda, \beta, \alpha_{1}, \alpha_{2} \) are to be tuned, since the parameters \( \bar{L}_{1}\) and \( \tilde{L}_{1}\) are usually not available. In order to overcome this obstacle, the adaptive 2-SMC controller blocks are coded.

\subsubsection{Adaptive 2-SMC blocks}

Assume that

(A5) There exists, at least locally, \emph{unknown} \( \bar{\tilde{L}}_{1}\) so that \( |\bar{\dot{\xi}}(x,t)|\leq \bar{\tilde{L}}_{1}\).

\subsubsection{Adaptive Super Twisting controller \cite{ref7}}  

is implemented and coded in the SMC Aero toolbox in equation (\ref{eq:21}) format with the adaptive gains

\begin{equation}\label{eq:24}
    \dot{\lambda} = 
     \begin{cases}
      \gamma sign(|\sigma|-\mu), \enspace if \: \lambda>\lambda_{m} \\
      \eta, \qquad \enspace \qquad \qquad \enspace \: if \: \lambda \leq \lambda_{m} \\
      \end{cases}, \enspace \beta = \epsilon \lambda
\end{equation}

where, \(\gamma > 0\) is an adaptation rate, \(\mu > 0\) is adaptation accuracy and \(\lambda_{m}, \epsilon, \eta\) are positive small constants.

\subsubsection{Adaptive Twisting controller block \cite{ref35}} is implemented and coded in the SMC Aero toolbox in equation (\ref{eq:22}) format with the adaptive gains

Control gain \(\propto \left( t \right)\) adaptation law is as follows

\vspace{-0.25cm}\begin{equation}\label{eq:25}
    \dot{\alpha_{1}} = 
     \begin{cases}
      \gamma_{1}\cdot sign(V(\sigma,\dot{\sigma})-\mu_{1}), \enspace if \: \alpha_{1}\geq \alpha_{1}min \\
      \eta_{1}, \qquad \enspace \qquad \qquad \enspace \qquad if \: \alpha_{1} < \alpha_{1}min \\
      \end{cases}, \enspace \alpha_{2} = 0.5\: \alpha_{1}
\end{equation}

where \( V(\sigma, \dot{\sigma}) \) is coded in a simplified format \( V(\sigma, \dot{\sigma}) = \sigma^{2} + c \cdot \dot{\sigma}^{2}, c > 0; \: \gamma_{1} > 0 \) is an adaptation rate, \( \mu_{1} > 0 \) is an adaptation accuracy and \(\alpha_{1\min},\eta_{1} > 0\ \)are small constants.

Note, that the adaptive STW and TW controllers in (\ref{eq:21}), (\ref{eq:22}), (\ref{eq:24}), and (\ref{eq:25}) provide non-overestimating control gains \cite{ref7,ref35}.

\begin{figure}[H]
    \centering
    \includegraphics[scale=1]{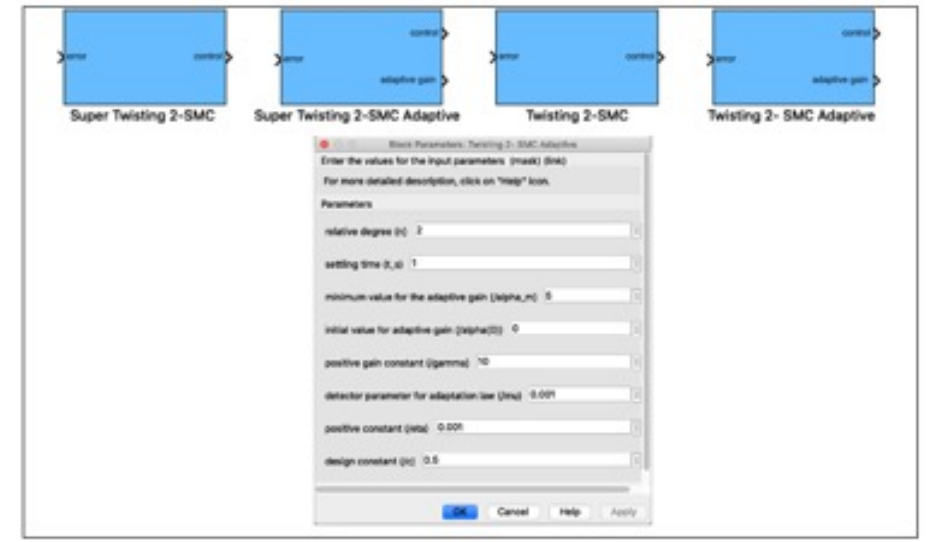}
    \caption{2-SMC library blocks with the Adaptive Twisting Controller block mask}
    \label{fig:fig10}
\end{figure}

The 2-SMC library blocks with the adaptive blocks mask are demonstrated in Figure~\ref{fig:fig10}.

\subsection{HOSMC Control Library}
This library consists of the blocks that address the SISO output tracking problem of relative degree \(r \geq 2\) for systems in equations (\ref{eq:2}) - (\ref{eq:5}) with \( m = 1\):

\begin{equation}\label{eq:26}
    e^{(r)} = \xi - v, \enspace u = G_{0}^{-1}(x)v 
\end{equation}

with \(e,\xi,v,u,G_{0} \in R^{1} \). The output tracking error \(e\) is employed as a sliding variable, i.e. \(\sigma = e\).

Assume that assumption (A1) holds for \(m = 1\). The discontinuous HOSM control that drives \(\sigma,\dot{\sigma},..\sigma^{r - 1} \rightarrow 0\) in finite time and keeps \(\sigma,\dot{\sigma}\ldots\sigma^{r - 1} = 0\) for all consecutive times is coded and implemented in the \emph{SMC Aero} toolbox in three formats:

\subsubsection{Quasi Continuous HOSM controller block [\cite{ref1}, \cite{ref5}]}

The Quasi Continuous HOSM controller block is implemented and coded for relative degrees \(r = 2,3,4,5\). For instance, the coded fifth order \(\left( r = 5 \right)\ \) Quasi Continuous HOSM controller is presented as

\begin{equation}\label{eq:27}
v = \ \alpha\frac{\left\lfloor \left. \ \sigma^{(4)} \right\rceil \right.\ ^{5} + {6\left\lfloor \left. \ \dddot{\sigma} \right\rceil \right.\ }^{\frac{5}{2}} + {5\left\lfloor \left. \ \ddot{\sigma} \right\rceil \right.\ }^{\frac{5}{3}}{+ 6\left\lfloor \left. \ \dot{\sigma} \right\rceil \right.\ }^{\frac{5}{4}} + \sigma}{\left| \sigma^{(4)} \right|^{5} + {6\left| \dddot{\sigma} \right|}^{\frac{5}{2}} + {5\left| \ddot{\sigma} \right|}^{\frac{5}{3}}{+ 6\left| \dot{\sigma} \right|}^{\frac{5}{4}} + \left| \sigma \right|}
\end{equation}

where \( \alpha > 0 \) is large enough. The following notation is used in (\ref{eq:27}):
\(\left\lfloor \left. \ \sigma \right\rceil \right.\ ^{\gamma} = \ \left| \sigma \right|^{\gamma}sign(\sigma)\), 
if \(\gamma > 0\) or \(\sigma \neq 0\);
\(\left\lfloor \left. \ \sigma \right\rceil \right.\ ^{0} = \ sign\ (\sigma)\).

\subsubsection{Nested HOSM controller block \cite{ref1,ref4}}

The \emph{Nested} HOSM controller block is implemented and coded for relative degrees \( r = 2,3,4,5 \). For instance, the coded fourth order \(\left( r = 4 \right)\ \)Nested HOSM controller is presented as

\begin{equation}\label{eq:28}
    v = \alpha \{\dddot{\sigma}+3(\ddot{\sigma}^{6}+\dot{\sigma}^{4}+|\sigma|^{3})^{\frac{1}{12}} \cdot sign [\ddot{\sigma}+(\dot{\sigma}^{4}+|\sigma|^{3})^{\frac{1}{6}} \cdot sign(\dot{\sigma}+0.5|\sigma|^{\frac{3}{4}}sign(\sigma))]\}
\end{equation}

\subsubsection{Adaptive Continuous HOSM controller block \cite{ref23,ref36}}

The \emph{Adaptive Continuous} HOSM controller block is implemented and coded for relative degrees \(r = 2,3,4,5\). For instance, the coded third order \(\left( r = 3 \right)\ \)\emph{Adaptive Continuous} HOSM controller is presented as

\begin{equation}\label{eq:29}
    v(t) = -v_{\sigma}(t)-v_{s}(t)
\end{equation}

where
\begin{equation}\label{eq:30}
    v_{\sigma}(\cdot) = \gamma_{1}|\sigma|^{\alpha_{1}}sign(\sigma)+\gamma_{1}|\dot{\sigma}|^{\alpha_{2}}sign(\dot{\sigma})+\gamma_{1}|\ddot{\sigma}|^{\alpha_{3}}sign(\ddot{\sigma})
\end{equation}

\begin{equation}\label{eq:31}
    v_{s}(t) = s(t)\dot{L}(t)/L(t)+\lambda(t)|s|^{\frac{1}{2}}sign(s)+\int_{0}^{t}\beta(\tau)sign((s(\tau)))d\tau
\end{equation}

where the auxiliary sliding variable \(s\) is defined as

\begin{equation}\label{eq:32}
    s(t) = \ddot{\sigma}(t)+\int_{0}^{t}v_{s}(\tau)d\tau
\end{equation}

The adaptive gains are defined as

\begin{equation}\label{eq:33}
    \lambda(t) = 2\sqrt{2\beta_{0}L(t)}
\end{equation}

\begin{equation}\label{eq:34}
\beta(t) = L(t)\beta_{0}    
\end{equation}

where \(\beta_{0} > 1\) is a fixed design scalar and the proposed adaptive element \(L(t)\) on which the gains \(\lambda(t)\) and \(\beta(t)\) depend is given by

\begin{equation}\label{eq:35}
    L(t) = l_{0}+l(t)
\end{equation}

where \(l_{0}\) is a (small) fixed positive (design) constant and

\begin{equation}\label{eq:36}
    \dot{l}(t) = -\rho(t)sign(\delta(t))
\end{equation}

where \(\delta\left(t \right)\) is defined as

\begin{equation}\label{eq:37}
    \delta(t) = L(t)-\frac{1}{\alpha\beta_{0}}|\bar{v}_{eq}(t)|
\end{equation}

\(\bar{v}_{eq}(t)\) is the approximation of the equivalent control, the scalar \(\alpha\) is chosen to satisfy \(0 < \alpha < \frac{1}{\beta_{0}} < 1\) and \(\epsilon\) is a small positive scalar chosen to ensure

\begin{equation}\label{eq:38}
\frac{1}{\alpha\beta_{0}}|\bar{v}_{eq}(t)|+\frac{\epsilon}{2}>\frac{1}{\beta_{0}}|v_{eq}(t)|    
\end{equation}

The design scalars \(\alpha\) and \(\epsilon\) represent the "safety margins"

The scalar in equation (\ref{eq:36}) is by definition

\begin{equation}\label{eq:39}
    \rho(t) = r_{0}+r(t)
\end{equation}

where \(r_{0}\) is a fixed positive design scalar and the time varying component \(r(t)\) satisfies

\begin{equation}\label{eq:40}
    \dot{r}(t) = 
    \begin{cases}
      \gamma|\delta(t)|, \enspace \: if \: |\delta(t)|> \delta_{0} \\
      0, \qquad \enspace \: \: otherwise.
    \end{cases} 
\end{equation}

where \(\delta_{0}\) is a (small) positive design scalar.

\begin{figure}[H]
    \centering
    \includegraphics[scale=0.8]{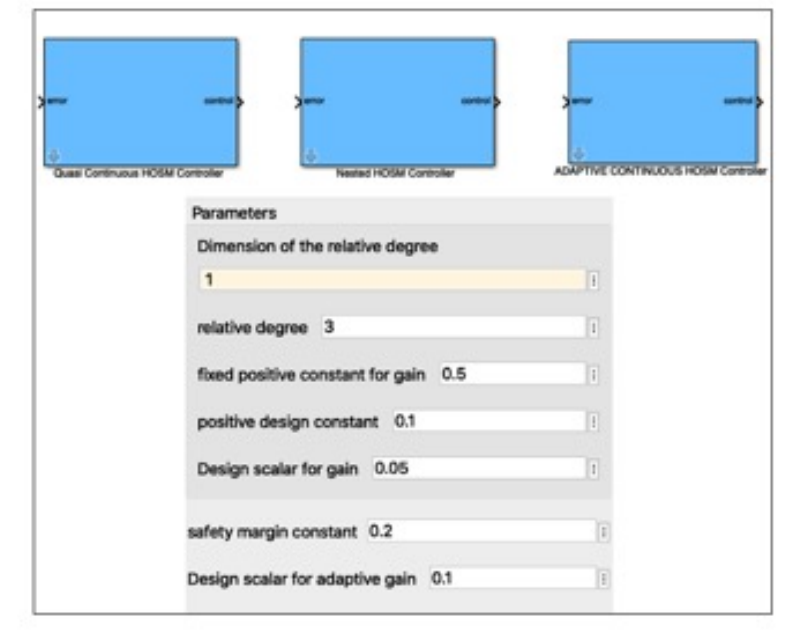}
    \caption{HOSM control library blocks with the Adaptive Continuous HOSM Controller mask}
    \label{fig:fig11}
\end{figure}

The HOSM control library blocks with a mask for the Continuous Adaptive HOSMC is given in Figure~\ref{fig:fig11}

\subsection{HOSM Differentiators Library \cite{ref1,ref4,ref37}}

This library consists of the blocks that address the problem of finding real-time robust estimations of \({\dot{f}}_{0}\left( t \right)\),\(\ \ddot{f_{0}}\left( t \right)\text{..\ }{f_{0}}^{k}\left( t \right)\). These differentiators were coded and implemented in the \emph{SMC Aero} toolbox in two formats: the differentiators that compute the exact derivative in the \emph{absence} of the measurement noise, and the differentiators that compute accurate derivatives in the \emph{presence} of the measurement noise.

\subsubsection{The robust HOSM differentiators} 

that compute the exact derivatives in the absence of measurement noise were coded for \( k = 1,2,3,4, 5\). Note that the parameters are available up to \(k = 12\). 

The HOSM differentiators are robust to bounded noises and feature optimal error asymptotic behavior \cite{ref37}.

For instance, the coded HOSM differentiator for \( k = 2\) is presented as

\begin{gather} \nonumber
    \dot{z}_{0} = -\lambda_{2} \bar{L}_{0}^{\frac{1}{3}}\left\lfloor \left. \ z_{0} - f(t) \right\rceil \right.\ ^{\frac{2}{3}}+z_{1} \\ \label{eq:41}
    \dot{z}_{1} = -\lambda_{1} \bar{L}_{0}^{\frac{2}{3}}\left\lfloor \left. \ z_{0} - f(t) \right\rceil \right.\ ^{\frac{1}{3}}+z_{2} \\ \nonumber
    \dot{z}_{2} = -\lambda_{0} \bar{L}_{0} sign(z_{0}-f(t)) 
\end{gather}

with the notation:
\(\left\lfloor \left. \ s \right\rceil \right.\ ^ {\gamma} = \ \left| s \right|^{\gamma}\text{sign}\ (s)\)
if \(\gamma > 0\) or \(s \neq 0\);
\(\left\lfloor \left. \ s \right\rceil \right.\ ^{0} = \text{\ sign}\ (s)\).

Here

\begin{itemize}
\item
  \(f(t) = f_{0}(t) +\chi(t) \) is the measured input function defined on \( t \in [0,\infty) \) consisting of a bounded Lebesgue-measurable noise \(\chi(t)\), and of an unknown base signal \(f_{0}\left( t \right)\), whose third derivative has a known Lipschitz constant \(\bar{L}_{0}>0\)
\item
  It is assumed here that \( |f_{0}^{(3)}| < \bar{L}_{0}\);
\item
  \( \lambda_{0} = 1.1, \lambda_{1} = 2.12, and \: \lambda_{2} = 3.0 \).
\end{itemize}

Note that the HOSM differentiator (\ref{eq:41}) provides the exact finite time convergence

\begin{equation}\label{eq:42}
 z_{0} \to f_{0}(t), z_{1} \to f_{0}^{(1)}(t), z_{2} \to f_{0}^{(2)}(t)
\end{equation}

in the absence of the measurement noise as soon as the second order sliding mode establishes in the differentiator/observer (\ref{eq:41}).

\subsubsection{HOSM Robust to Noise Filtering Differentiators}
that compute the accurate derivatives in the presence of the measurement noise were coded for \(k = 1,2,3,4,5\). The parameters are available up to \(k = 12\)

Note that the differentiator (\ref{eq:41}) is a particular case of the HOSM Filtering Differentiators that compute exact derivatives in the absence of noises, feature the optimal error asymptotic behavior in the presence of bounded noises, and are capable of attenuating very large noises with bounded iterated integrals \cite{ref37}.

Here the order of the differentiator \(k\) is presented as \( k = n_{d}+n_{f} \), where \(n_{d}\) is called the \emph{differentiation order and} \(n_{f} \geq 0 \) is called the \emph{filtering order.}

It is assumed that

(A5) \(|f_{0}^{(n_{d}+1)}| \leq \bar{L}_{0}\)

For instance, differentiator (\ref{eq:41}) has the differentiation order \(n_{d} = 2\) and the filtering order \(n_{f} = 0\). The coded filtering differentiator for \( k = 5, n_{d} = 3, n_{f} = 2 \) has the form

\begin{gather} \nonumber \nonumber
\dot{w}_{1} = -\lambda_{5} \bar{L}_{0}^{\frac{1}{6}}\left\lfloor \left. \ w_{1} \right\rceil \right.\ ^{\frac{5}{6}}+w_{2} \\ \nonumber
\dot{w}_{2} = -\lambda_{4} \bar{L}_{0}^{\frac{2}{6}}\left\lfloor \left. \ w_{1} \right\rceil \right.\ ^{\frac{4}{6}}+z_{0}-f(t) \\ \nonumber
\dot{z}_{0} = -\lambda_{3} \bar{L}_{0}^{\frac{3}{6}}\left\lfloor \left. \ w_{1} \right\rceil \right.\ ^{\frac{3}{6}}+z_{1} \\ \label{eq:43}
\dot{z}_{1} = -\lambda_{2} \bar{L}_{0}^{\frac{4}{6}}\left\lfloor \left. \ w_{1} \right\rceil \right.\ ^{\frac{2}{6}}+z_{2} \\\nonumber
\dot{z}_{2} = -\lambda_{1} \bar{L}_{0}^{\frac{5}{6}}\left\lfloor \left. \ w_{1} \right\rceil \right.\ ^{\frac{1}{6}}+z_{3} \\\nonumber
\dot{z}_{3} = -\lambda_{0} \bar{L}_{0} sign (w_{1}) \nonumber
\end{gather}

where

\begin{itemize}
\item
  \(f(t) = f_{0}(t) +\chi(t) \) is the measured input function defined on \( t \in [0,\infty) \) consisting of a bounded Lebesgue-measurable noise \(\chi(t)\), and of an unknown base signal \(f_{0}\left( t \right)\), whose fourth derivative has a known Lipschitz constant \(\bar{L}_{0}>0\);
\item
  In accordance with assumption (A5) inequality  \(|f_{0}^{(4)}| \leq \bar{L}_{0}\) holds;
\item
  Parameters \(\lambda_{0},\lambda_{1}\ldots\lambda_{5}\) for \(n = 5\) are 1.1, 6.75, 20.26, 32.24, 23.72, and 7 respectively.
\end{itemize}

Note that

\(z_{0} \rightarrow f_{o}\left( t \right)\),
\(z_{1} \rightarrow {f_{o}}^{\left( 1 \right)}(t)\),
\(z_{2} \rightarrow {f_{o}}^{\left( 2 \right)}(t)\),
\(z_{3} \rightarrow {f_{o}}^{\left( 3 \right)}(t)\) exactly in finite time in the absence of measurement noise as soon as the second order sliding modes establish in the differentiator/observer (\ref{eq:43}).

\begin{figure}[H]
    \centering
    \includegraphics[scale=0.8]{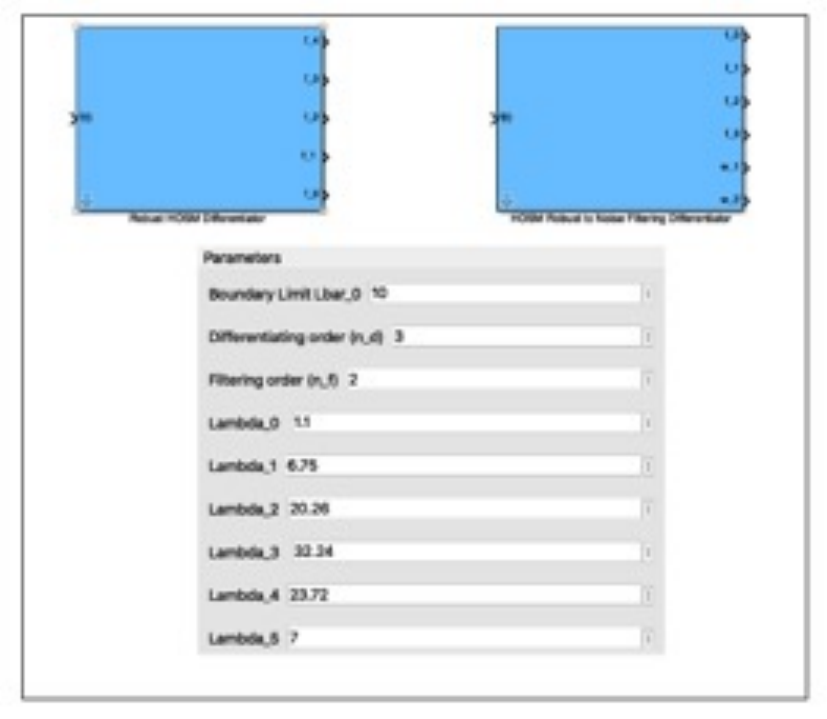}
    \caption{HOSM Differentiators library blocks with a mask for HOSM Robust to Noise Filtering Differentiator block}
    \label{fig:fig12}
\end{figure}

The HOSM Differentiators library blocks with a mask for the Filtering HOSM Differentiator block are shown in Figure~\ref{fig:fig12}.

\subsection{SMC for Lunar Lander Library}

SMC for Lunar Lander (RPL) library was designed to feature a specific "off-pulsing" control property, i.e. to generate control pulses with a given duration. A pulse-width modulation (PWM) techniques is employed to implement this SMC RPL feature.

The modified sliding variable (\( \bar{\sigma} \)) is defined as a combination of sliding variable \((\sigma\)) and dither signal \((\Delta t\))

\begin{equation}\label{eq:44}
    \bar{\sigma} = \sigma + \Delta t, \enspace \Delta t = a \sin (\omega t) \enspace \& \enspace \omega = \frac{2\pi}{N}
\end{equation}

where \(a,\ N,\omega\ \)are amplitude, period and frequency of dither signal respectively, which will be chosen by the user.

The modified signum function (\(\overline{sign}\)) is defined as a combination of signum function (\(sign\)) of varying dead band and fixed duration
of the pulse (\(S\& H\)).

The PWM of the sliding variable carries the following characteristics:

\emph{The Time Varying Dead band} (\(\delta T\)): The time varying dead band is enforced to obtain sparse control pulses at the beginning of the controlled RPL landing, and then control pulses become denser towards the end in order to pinpoint the soft landing.

\emph{The time period} (\(T\)) is defined as the time interval for controlled landing, further \(T\) is divided into \(n\) time intervals (\(t_{1},\ t_{2},t_{3}\ldots.,t_{n}\)), and \(n\) dead bands \(\left( \delta_{1},\delta_{2}\ldots.,\delta_{n} \right)\) are allocated to each of these time intervals. The larger dead bands are allocated at the beginning of the controlled descent, and then decrease closer to the landing moment.

\emph{Sample and Hold} (\(S\&H\)): A frequency of firing of the thrusters is regulated using PWM technique with the varying dead band, and S\&H block in MATLAB/Simulink libraries.

SMC off- pulse controller is now computed in a subroutine

\begin{equation}\label{eq:45}
    v = PWM(\rho \cdot \overline{sign}(\bar{\sigma}), \delta T, \omega, a)
\end{equation}

\begin{figure}[H]
    \centering
    \includegraphics[scale=1]{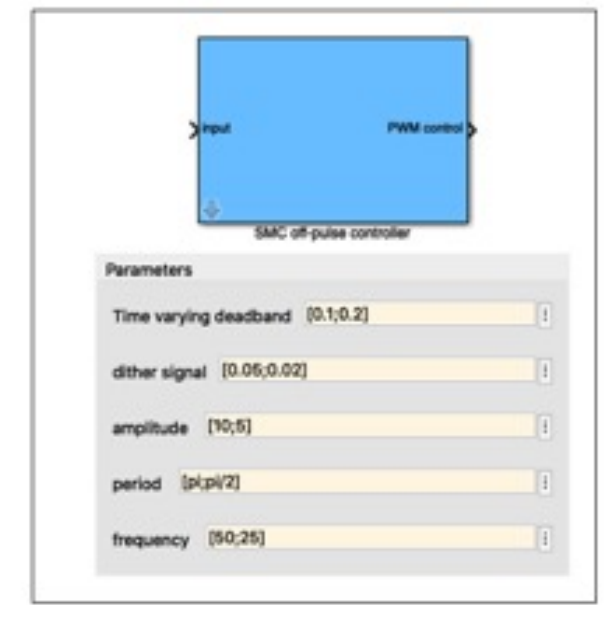}
    \caption{SMC off-pulse controller with a mask overview}
    \label{fig:fig13}
\end{figure}

The SMC off-pulse controller block with a mask is shown in Figure~\ref{fig:fig13}.

Note that

\begin{itemize}
\item
  off-pulsing control with given duration of the pulse is achieved in (\ref{eq:45}) using PWM of sliding variable connected to \emph{S\&H} block, which is needed to enforce a duty cycle (frequency) of pulses and their durations.
\item
  The efficacy of \emph{adaptive} 1-SMC with time varying dead band is verified on a case study of RPL the soft descent control design and simulation
\end{itemize}

\subsubsection{ Architecture of SMC for Lunar Lander}

Sliding mode control for Lunar Lander library mainly comprised of three blocks namely

\begin{enumerate}
\def\labelenumi{\arabic{enumi})}
\item
  1-SMC Lander
\item
  2-SMC Lander and
\item
  HOSMC Lander
\end{enumerate}

\begin{figure}[H]
    \centering
    \includegraphics[scale=1]{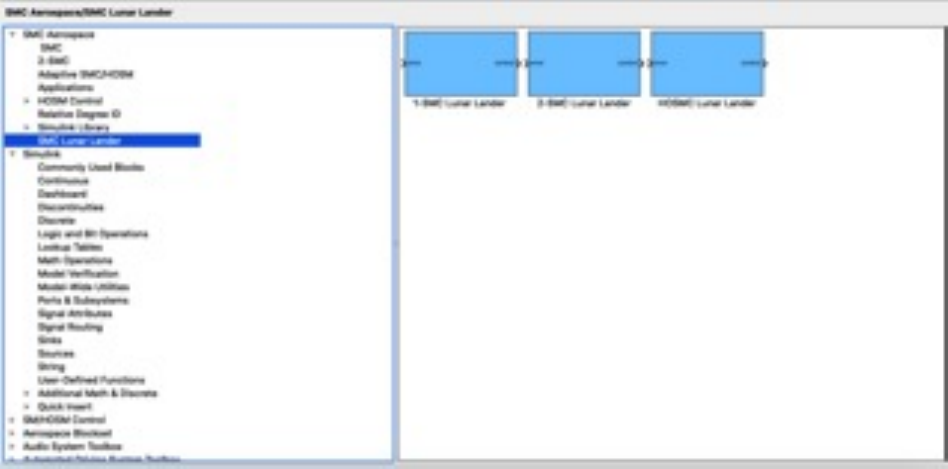}
    \caption{Proposed structure of SMC for Lunar Lander}
    \label{fig:fig14}
\end{figure}

A proposed architecture/structure of the SMC for lunar lander is presented in Figure~\ref{fig:fig14}.

\subsubsection{Architecture of SMC for Launch Vehicle}

Sliding mode control for Launch vehicle library mainly comprised of three blocks namely

\begin{enumerate}
\def\labelenumi{\arabic{enumi})}
\item
  1-SMC launch vehicle
\item
  2-SMC launch vehicle and
\item
  HOSMC launch vehicle
\end{enumerate}

\begin{figure}[H]
    \centering
    \includegraphics[scale=1]{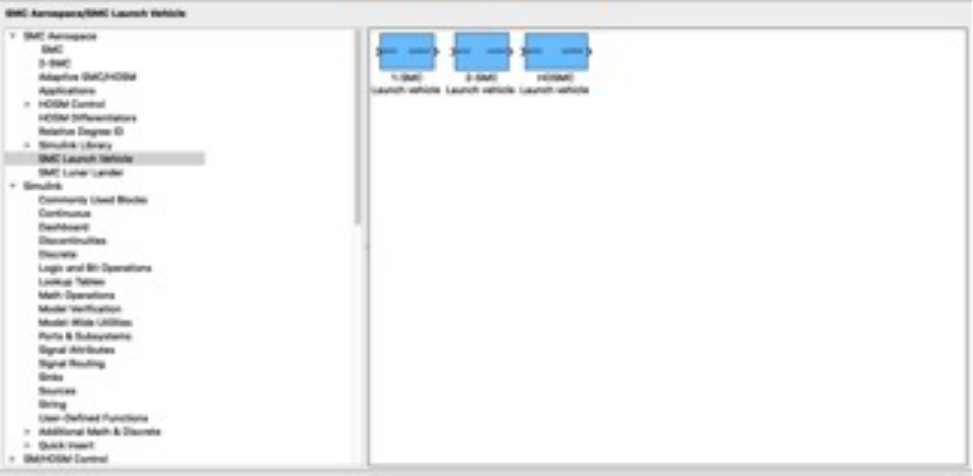}
    \caption{Proposed structure of SMC for Launch Vehicle}
    \label{fig:fig15}
\end{figure}

A proposed architecture/structure of the SMC for launch vehicle is presented in Figure~\ref{fig:fig15}.

\section{First Case Study: Resource Prospector Lander Control and Simulation \cite{ref43}}

A purpose of this section is to demonstrate the capabilities of the developed \emph{SMC Aero} toolbox on the adaptive 1-SMC and 2-SMC design and simulation of the RPL (Figure~\ref{fig:fig1}) soft landing on the moon.

A simplified mathematical pitch plane model of RPL in descent is taken as \cite{ref8,ref29}

\begin{equation}\label{eq:46}
    \ddot{\chi} = A\chi + Bu + g + \varphi
\end{equation}

where, 
\(\chi = [\theta, x]^{T} \in R^{2}\) is a vector-output, \(A = 0 \in R^{2}\) is a state coefficient matrix and \(B =\)  \(\begin{bmatrix} \frac{l_{a}}{J_{yy}} & 0 \\[0.3cm] \frac{sin\theta}{m} & \frac{sin\theta}{m} \end{bmatrix}\) \( \in R^{2x2}\) is a control coefficient matrix, \(u = \left\lbrack u_{a},u_{d} \right\rbrack^{T}\in R^{2}\) is a control vector, \(g = \lbrack{0, - g_{m}\rbrack}^{T}\in R^{2}\) is a gravity coefficient matrix and \(\varphi = \lbrack{\varphi_{a}\left(t\right),\varphi_{d}(t)\rbrack}^{T}\in R^{2}\) is a perturbation vector.

The parameters that contribute to the RPL mathematical modeling are presented as \(\theta\left( \deg \right)\ \)is an RPL attitude (pitch angle), and \(x(m)\) is the RPL altitude; \(g_{m} = 1.6\ (m/s^{2})\) is a lunar gravitational constant, \(l_{a} = 2.8\ \left( m \right)\text{\ \ }\)is a distance between the
attitude control force and a center of gravity, \(J_{\text{yy}} = 300\ (kg.m/s^{2}),\) \(m = 1000\ \left( \text{kg} \right)\ \)is the RPL moment of inertia
(assumed constant), is the RPL mass (assumed constant); \( u_{a}, u_{d} (N)\) are attitude and descent control forces respectively, \(\varphi_{a}(t)(rad/s^{2})\ \) and \(\varphi_{d}(t)(m/s^{2})\ \)are bounded perturbations that are due to modeling errors and uncertainties (due to vacuum flight, external disturbances are negligible during lunar descent).

The control problem here is in designing \emph{off-pulse control functions} in terms of \(u_{a},u_{d}\ \) with the pulse duration not less than \(50\ ms\) that drive the tracking errors \(e_{\theta} = \theta_{c} - \theta\) and \(e_{x} = x_{c} - x\) to the domains \(\left| e_{\theta} \right| \leq \varepsilon_{\theta},\ \ \left| e_{x} \right| \leq \varepsilon_{x}\ \) where \(\varepsilon_{\theta},\varepsilon_{x} > 0\ \)by the landing time \(T\), where \(\theta_{c}\) and \(x_{c}\) are command trajectories that guarantee soft landing if followed. To minimize the effects on the attitude stabilization the control \(u_{d}\) is to be allocated symmetrically as \(u_{d}\text{\ left\ }\)and \(u_{d}\text{\ right}\)

\begin{equation}\label{eq:47}
       u_{d}left = u_{d}right = \frac{u_{d}}{2} 
\end{equation}

Note that the RPL (Figure~\ref{fig:fig1}) is controlled in descent by twelve attitude and twelve descent control thrusters \cite{ref8,ref29} that generate pitch and
yaw control forces/torques. Therefore, the attitude and descent control functions \(u_{a}, u_{d}\) are to be allocated to the thrusters. A control allocation is not done in this simplified study, whose goal is to demonstrate capabilities of the \emph{SMC Aero} toolbox to aerospace vehicle sliding mode control design and simulation.

\subsection{RPL SMC Design Using SMC Aero toolbox}

\emph{Step 1:} Practical relative degree identification

The PRD ID block is used to identify the vector practical degree \(r = \left\lbrack r_{\theta},\ r_{x} \right\rbrack = \lbrack 2,2\rbrack\). The corresponding MATLAB editor code of a performance analyzer and PRD ID block with embedded mask feature are shown in Figures \ref{fig:fig4} and \ref{fig:fig5}.

\emph{Step 2:} Sliding variable design

Given \(m = 2\) and the desired settling times \(t_{\text{sa}} = \ 2\ sec,\ \ t_{\text{sx}} = \ 5\ sec\ \ \) in the attitude and descent sliding modes, the sliding variables \(\sigma_{a},\ \sigma_{d}\) are designed in a form (\ref{eq:9}) using the block Sliding Variable Design, whose structure is presented in Figure~\ref{fig:fig8}.

As a result, the following attitude and descent sliding variables are obtained:

\begin{equation}\label{eq:48}
\sigma_{a}  = {\dot{e}}_{a} + 7e_{a} + 25\int e_{a}dt
\end{equation}

\begin{equation}\label{eq:49}
\sigma_{d}  = {\dot{e}}_{d} + 2.8e_{d} + 4\int e_{d}dt
\end{equation}

where \(e_{\theta} = \theta_{c} - \theta,e_{x} = x_{c} - x\) with \(\theta_{c}, x_{c}\) to be attitude and altitude commanded trajectories in the soft-landing mode.

\emph{Step 3:} Adaptive 1-SMC design with varying dead band and limited time duration for thrust firing.

The virtual control functions \(v = [v_{a}, v_{d}]^{T} \in R^{2} \) that are to be designed are introduced in accordance with equation \ref{eq:13} and computed as

\begin{equation}\label{eq:50}
    v = Bu = 10^{-3} \begin{bmatrix}
          9.3 & 0 \\ sin\theta & sin\theta 
    \end{bmatrix} \begin{bmatrix}
          u_{a} \\ u_{d}
    \end{bmatrix}
\end{equation}

As soon as virtual control vector are designed, the original controls are computed as

\begin{equation}\label{eq:51}
    u = B^{-1}v = 10^{3} \begin{bmatrix}
          \frac{1}{9.3} & 0 \\[0.3cm] \frac{1}{sin\theta} & \frac{1}{sin\theta}
    \end{bmatrix} \begin{bmatrix}
          v_{a} \\ v_{d}
    \end{bmatrix}
\end{equation}

\begin{figure}[H]
    \centering
    \includegraphics[scale=1]{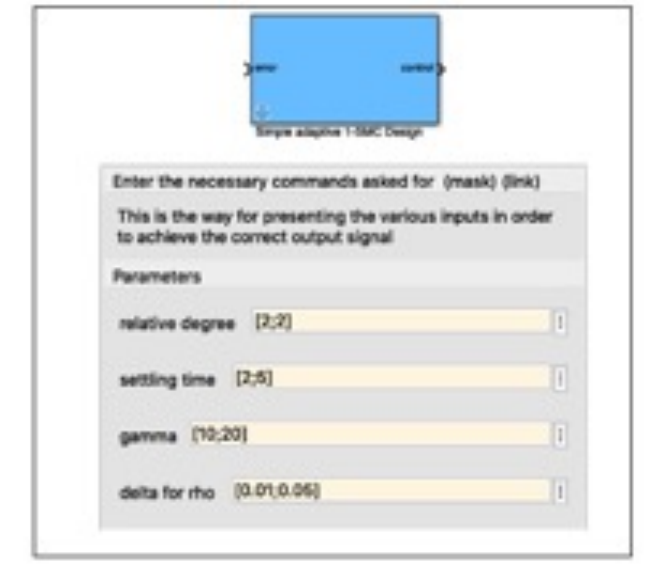}
    \caption{Adaptive 1-SMC Design block with the mask overview}
    \label{fig:fig16}
\end{figure}

Two adaptive 1-SMCs are designed using the adaptive algorithm in equation (\ref{eq:15}) individually for \(v_{a}\ \)and \(v_{d}\) presented in a block Adaptive 1-SMC Design, whose structure and mask overview are presented in Figure~\ref{fig:fig16}.

\emph{Step 4:} The implementation of the adaptive 1-SMC controllers in the "\emph{SMC Aero}" toolbox environment using a \emph{virtual} control design:

\begin{equation}\label{eq:52}
    v_{a} = \rho_{a} sign(\sigma_{a}), \enspace v_{d} = \rho_{d} sign(\sigma_{d})
\end{equation}

Adaptive attitude and descent control gains \(\rho_{a}{,\ \rho}_{d}\) are employed using the adaptation law is presented as

\begin{equation}\label{eq:53}
\begin{aligned}
    \dot{\rho_{a}}(t) = 
    \begin{cases}
      \gamma_{a}|\sigma_{a}(t)|, \enspace \enspace if \: |\sigma_{a}(t)| > \delta_{a} \\
      0, \qquad \qquad \: otherwise.
    \end{cases}  \\
     \dot{\rho_{d}}(t) = 
    \begin{cases}
      \gamma_{d}|\sigma_{d}(t)|, \enspace \enspace if \: |\sigma_{d}(t)| > \delta_{d} \\
      0, \qquad \qquad \: otherwise.
    \end{cases} 
\end{aligned}
\end{equation}

where \(\delta_{a},\ \delta_{d} > 0\) are the parameters of the attitude and descent adaptation accuracy respectively.

Finally, the adaptive 1-SMC controllers are designed as

\begin{equation}\label{eq:54}
    u_{a} = 107.5v_{a}, u_{d} = -107.5v_{a}+\frac{10^{3}}{sin\theta}v_{d}
\end{equation}

where the control gains are adapted in accordance with equation \ref{eq:53}.

Note that derived adaptive SMCs of varying dead band are to be implemented using PWM and S\(\&\)H blocks.

\subsection{Simulation Setup}

\begin{itemize}
\item
  The simulated command trajectory approximates the guidance profile required for soft landing on the moon, are defined as \cite{ref29}
\end{itemize}

\begin{gather}\label{eq:55}
    \theta_{c} = 90(deg) \\ \nonumber
    x_{c} = (6.9 \times 10^{-4})t^{3} - 0.23t^{2}-10t+6000(m)
\end{gather}

\begin{itemize}
\item
  The controlled landing duration is given as
\item
  Amplitude and frequency of dither signal in PWM block are chosen as \(a = 0.05\ \)and \(\omega = 20\ Hz\) for both attitude and descent
  case, and the Sample and Hold block duration is selected as \(0.05 s\).
\item
  The perturbations
\end{itemize}

\begin{gather}\label{eq:56}
    \varphi_{a}(t) = 0.06sin(0.2t)(rad/s^{2}) \\ \nonumber
    \varphi_{d}(t) = 0.5sin(0.2t)(rad/s^{2})
\end{gather}

\begin{quote}
that are due to misalignment of control forces and model uncertainties are used for the simulation purpose only to demonstrate the robustness of adaptive 1-SMC controllers;
\end{quote}

\begin{itemize}
\item
  The initial conditions:

\begin{gather*}
    \theta(0) = 91.67 deg, \enspace \dot{\theta}(0) = 5.73 (deg/s) \\
    x(0) = 6500 (m), \enspace \dot{x}(0) = -10 (m/s)
\end{gather*}

\item
  The control limits:
\end{itemize}

\[\left| u_{a} \right| \leq \ 15\ \left( N \right),\ 0 \leq \left| u_{d} \right| \leq \ 2800\ (N)\]

\begin{itemize}
\item
  Time intervals and corresponding time varying dead bands \(\delta_{ia}, \delta_{id}, \: i = 1,2,3\) for Attitude and Descent cases respectively are presented in the Table~\ref{table:1}
\end{itemize}

\begin{table}[h!]
\centering
\caption{Time intervals and corresponding time varying dead bands.}
\vspace{0.3cm}\begin{tabular}{c c c c} 
 \hline
 Time interval (s) & 0-100 & 100-200 & 200-240 \\ [0.5ex] 
 \hline
 Attitude dead band \(\delta_{ia}, i = 1,2,3\) & 0.05 & 0.025 & 0.01 \\ 
 Descent dead band \(\delta_{id}, i = 1,2,3\)  & 0.1 & 0.075 & 0.05 \\ 
 \hline
\end{tabular} 
\label{table:1}
\end{table}

\subsection{Simulation Diagram}

\begin{figure}[H]
    \centering
    \includegraphics[scale=0.8]{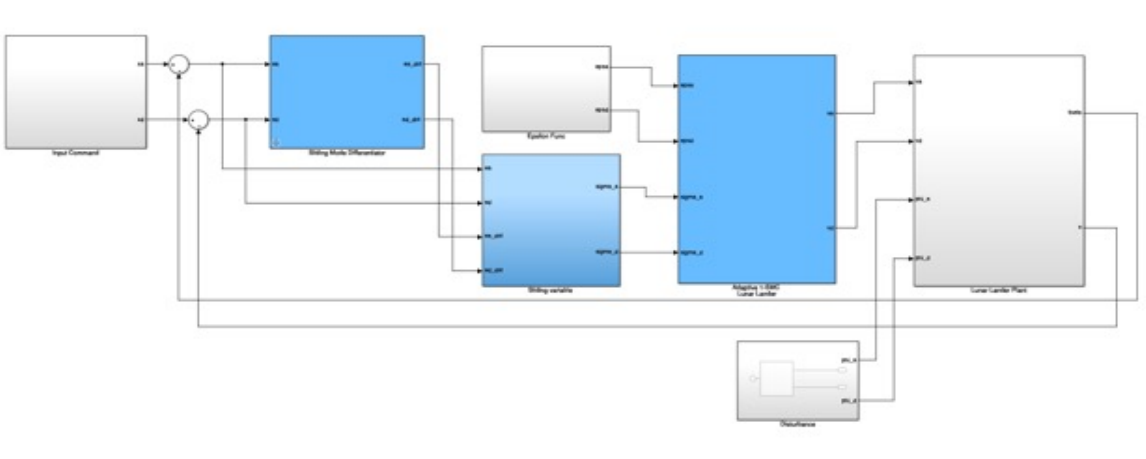}
    \caption{Simulation diagram of an RPL controlled by adaptive 1-SMC controllers}
    \label{fig:fig17}
\end{figure}

A simulation diagram of an RPL controlled by the adaptive 1-SMC controllers in (\ref{eq:53}, \ref{eq:54}) is presented in the Figure~\ref{fig:fig17}.

\subsection{Simulation Results and Discussions}\label{simulation results and discussions}

\begin{figure}[H]
    \centering
    \includegraphics[scale=0.25]{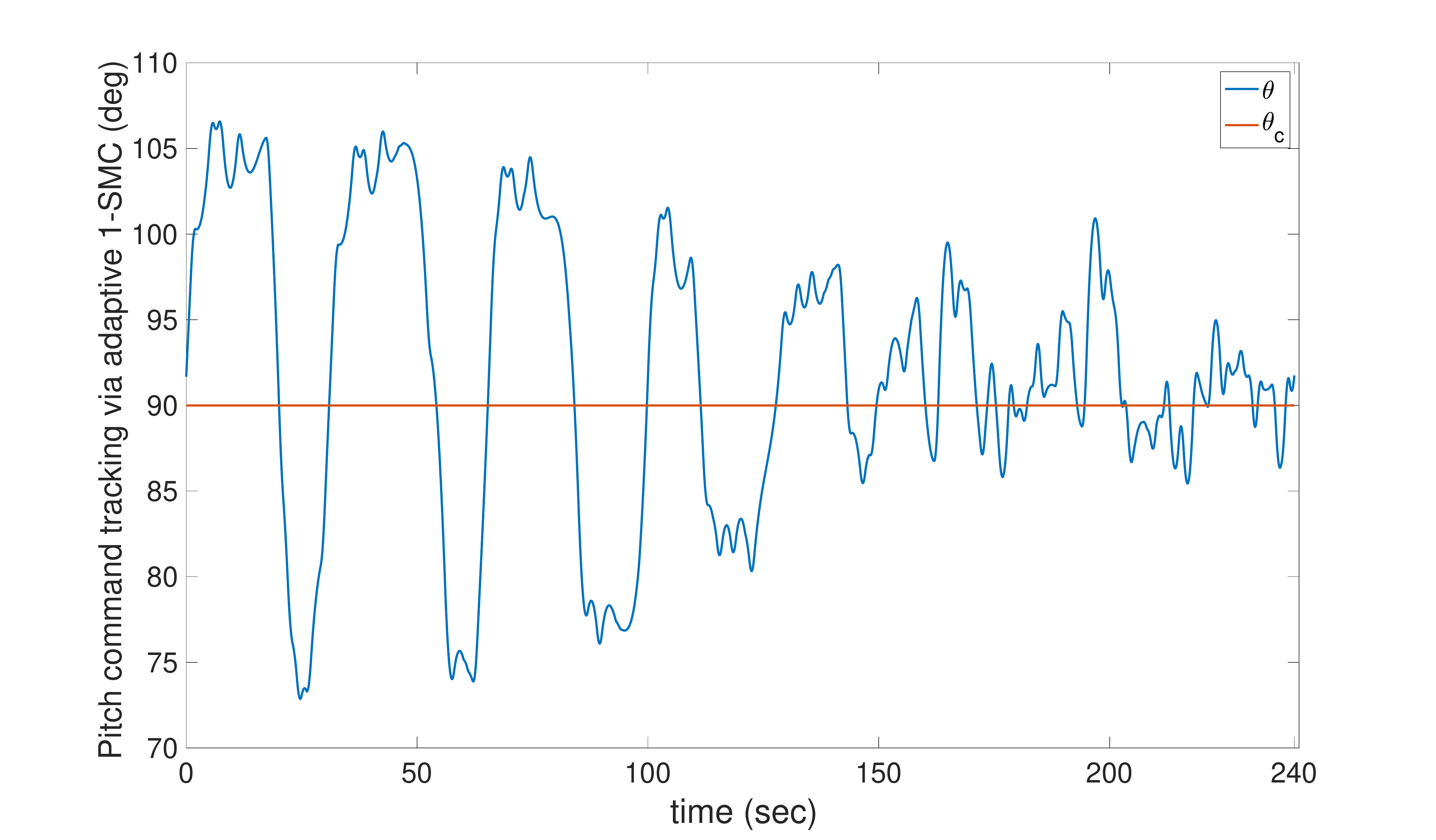}
    \caption{Pitch command tracking via adaptive 1-SMC}
    \label{fig:fig18}
\end{figure}

\begin{figure}[H]
    \centering
    \includegraphics[scale=0.25]{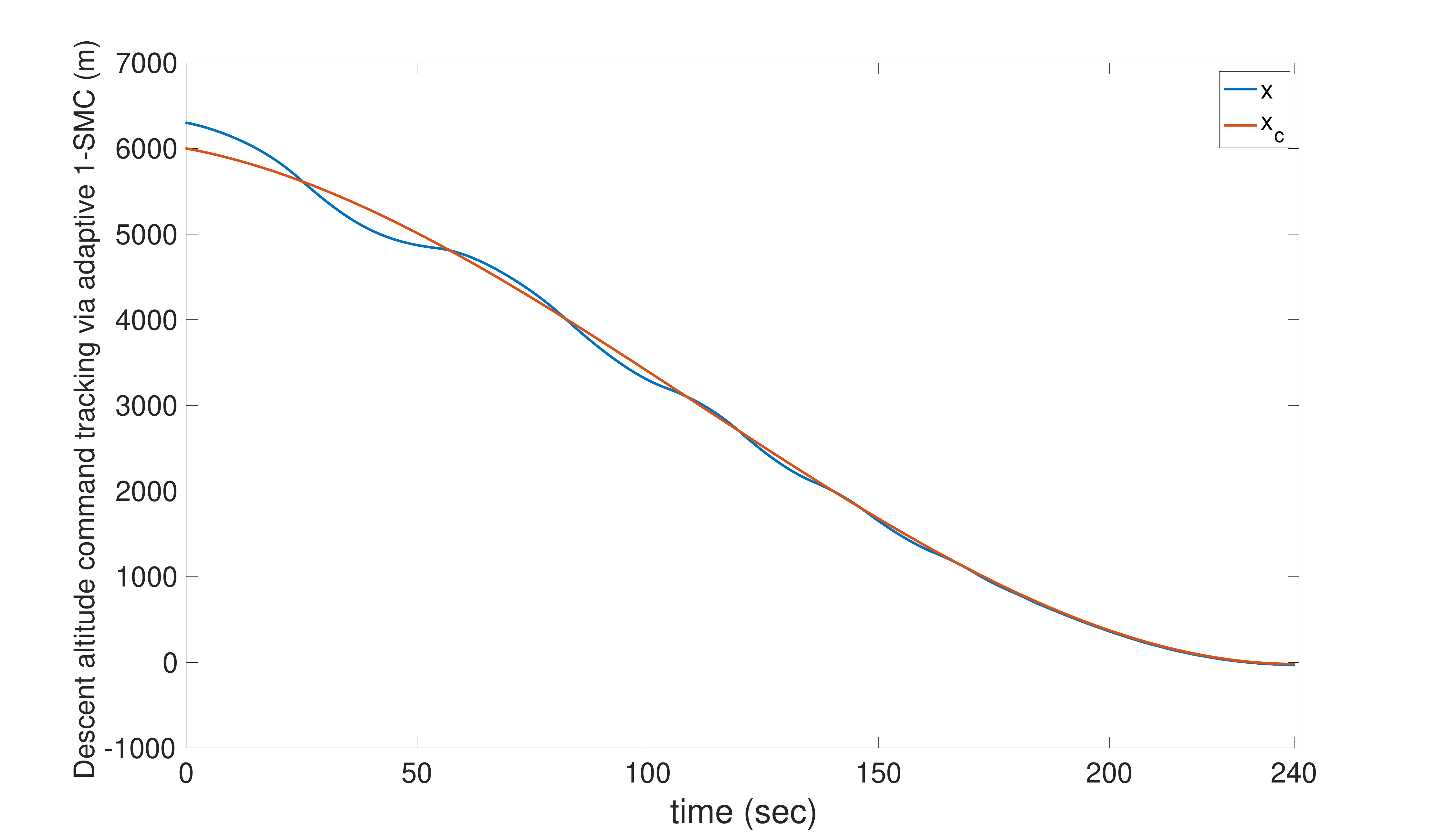}
    \caption{Descent altitude command tracking via adaptive 1-SMC}
    \label{fig:fig19}
\end{figure}

\begin{figure}[H]
    \centering
    \includegraphics[scale=0.25]{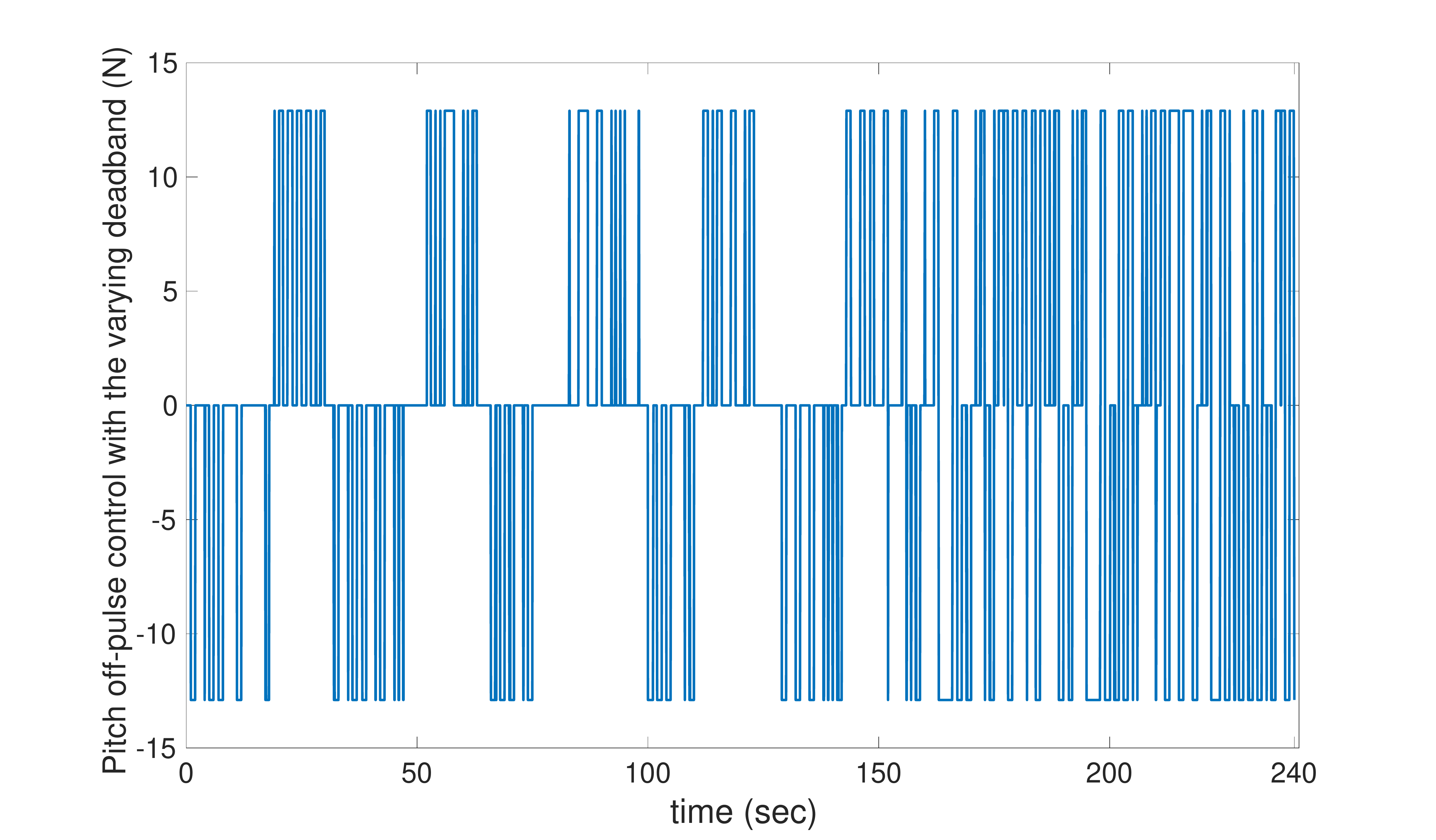}
    \caption{Pitch off-pulse control with the varying dead band}
    \label{fig:fig20}
\end{figure}

\begin{figure}[H]
    \centering
    \includegraphics[scale=0.25]{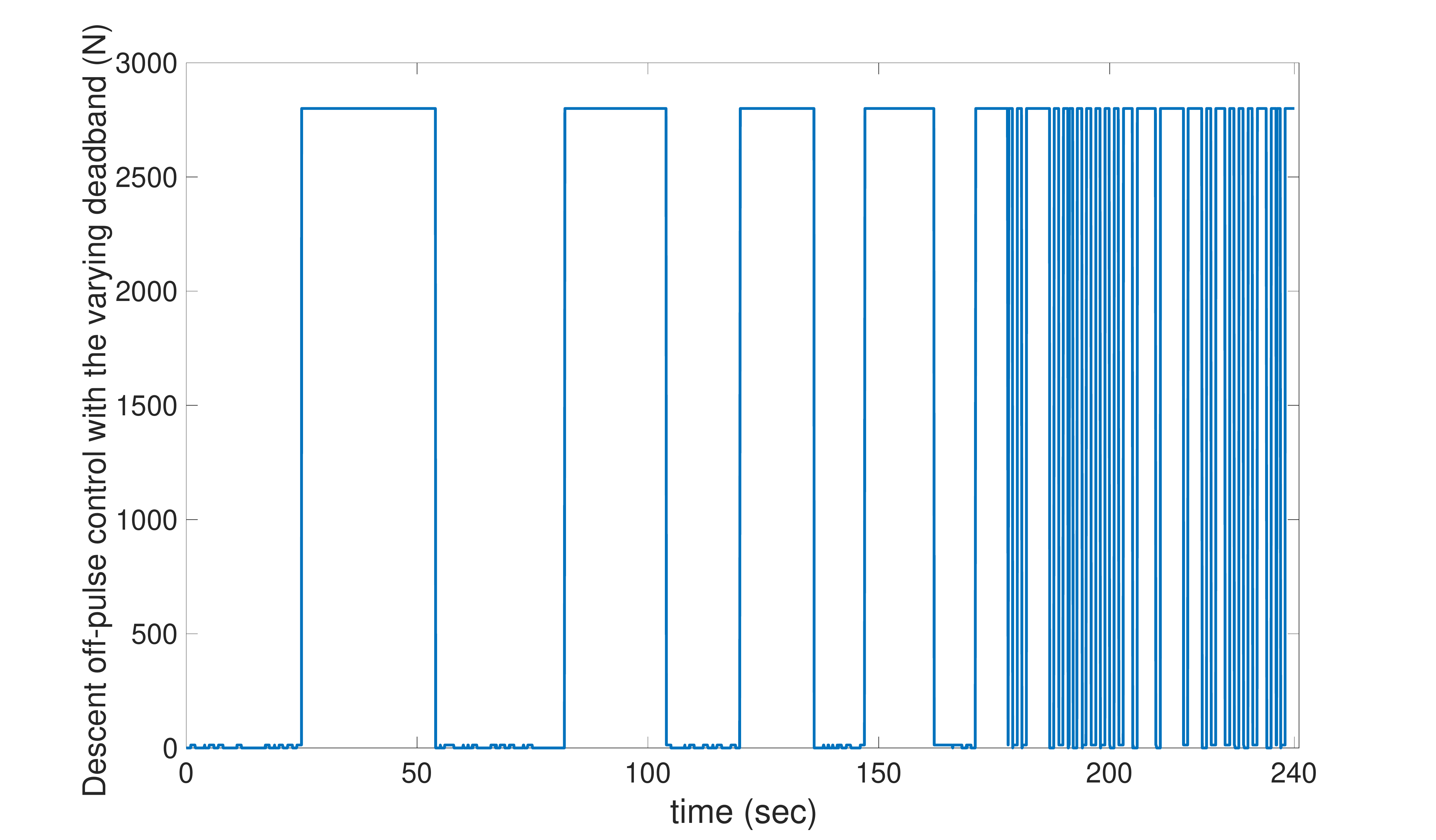}
    \caption{Descent off-pulse control with the varying dead band}
    \label{fig:fig21}
\end{figure}

\begin{figure}[H]
    \centering
    \includegraphics[scale=0.25]{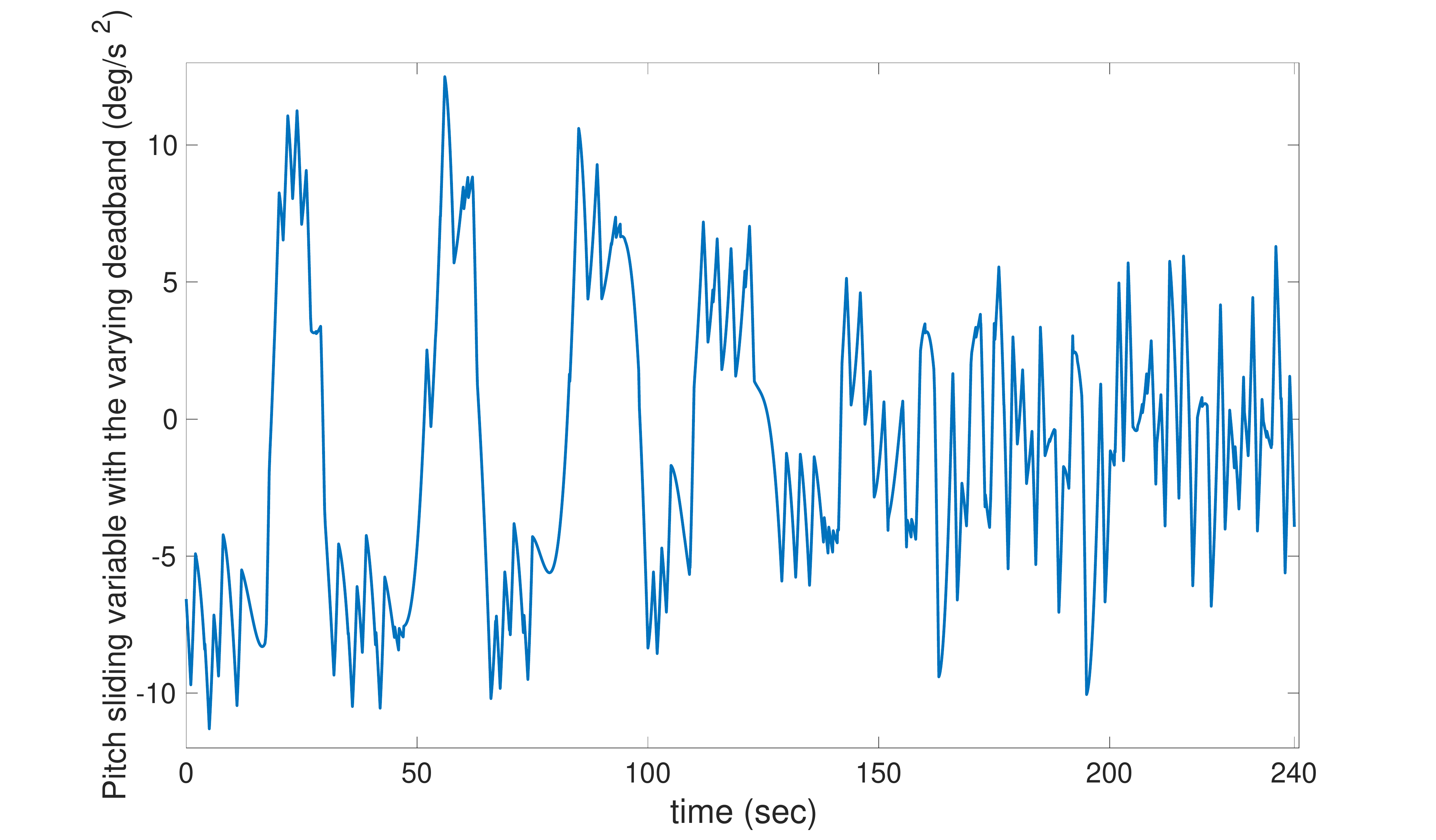}
    \caption{Pitch sliding variable with the varying dead band}
    \label{fig:fig22}
\end{figure}

\begin{figure}[H]
    \centering
    \includegraphics[scale=0.25]{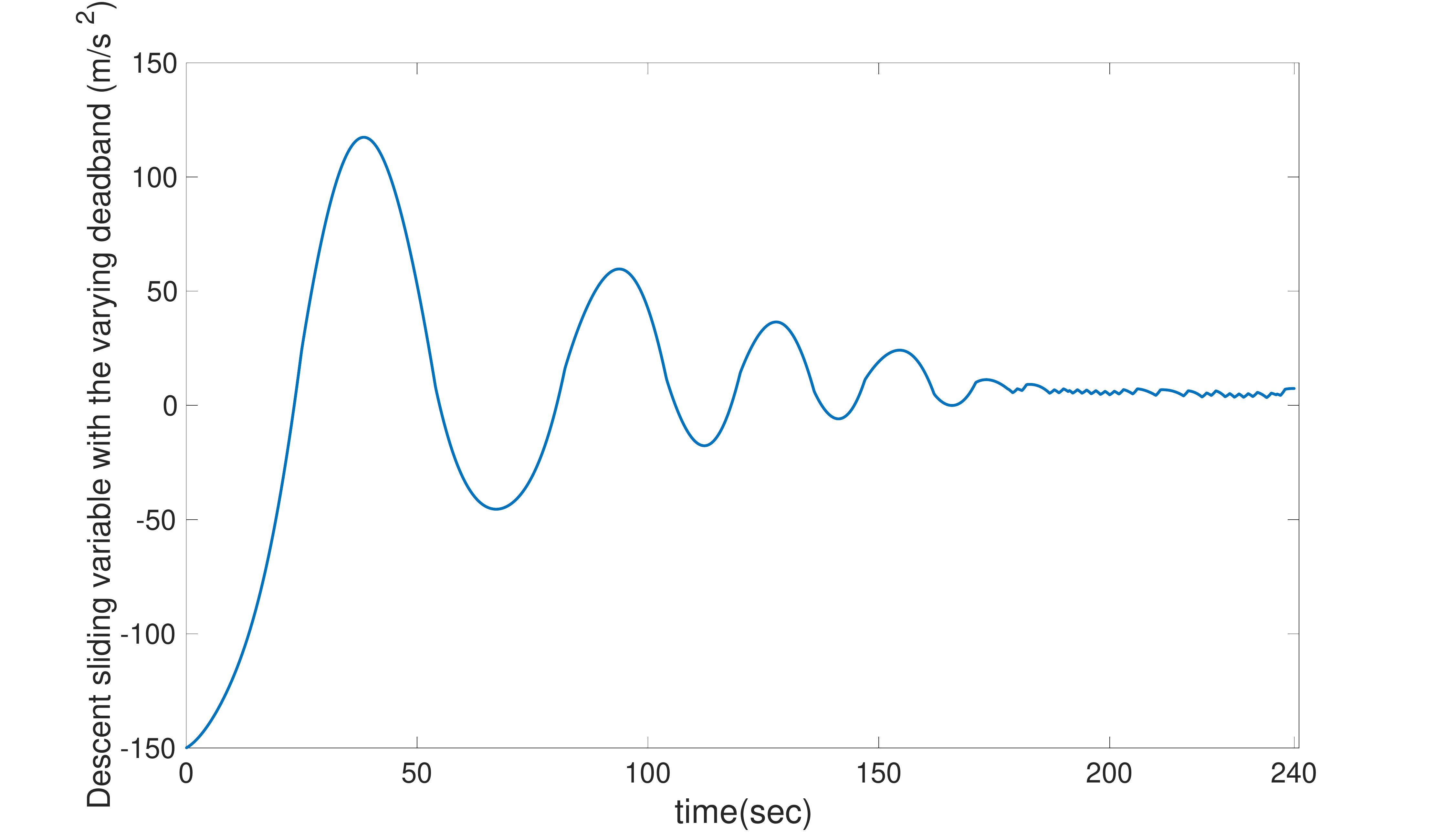}
    \caption{Descent sliding variable with the varying dead band}
    \label{fig:fig23}
\end{figure}

\begin{figure}[H]
    \centering
    \includegraphics[scale=0.25]{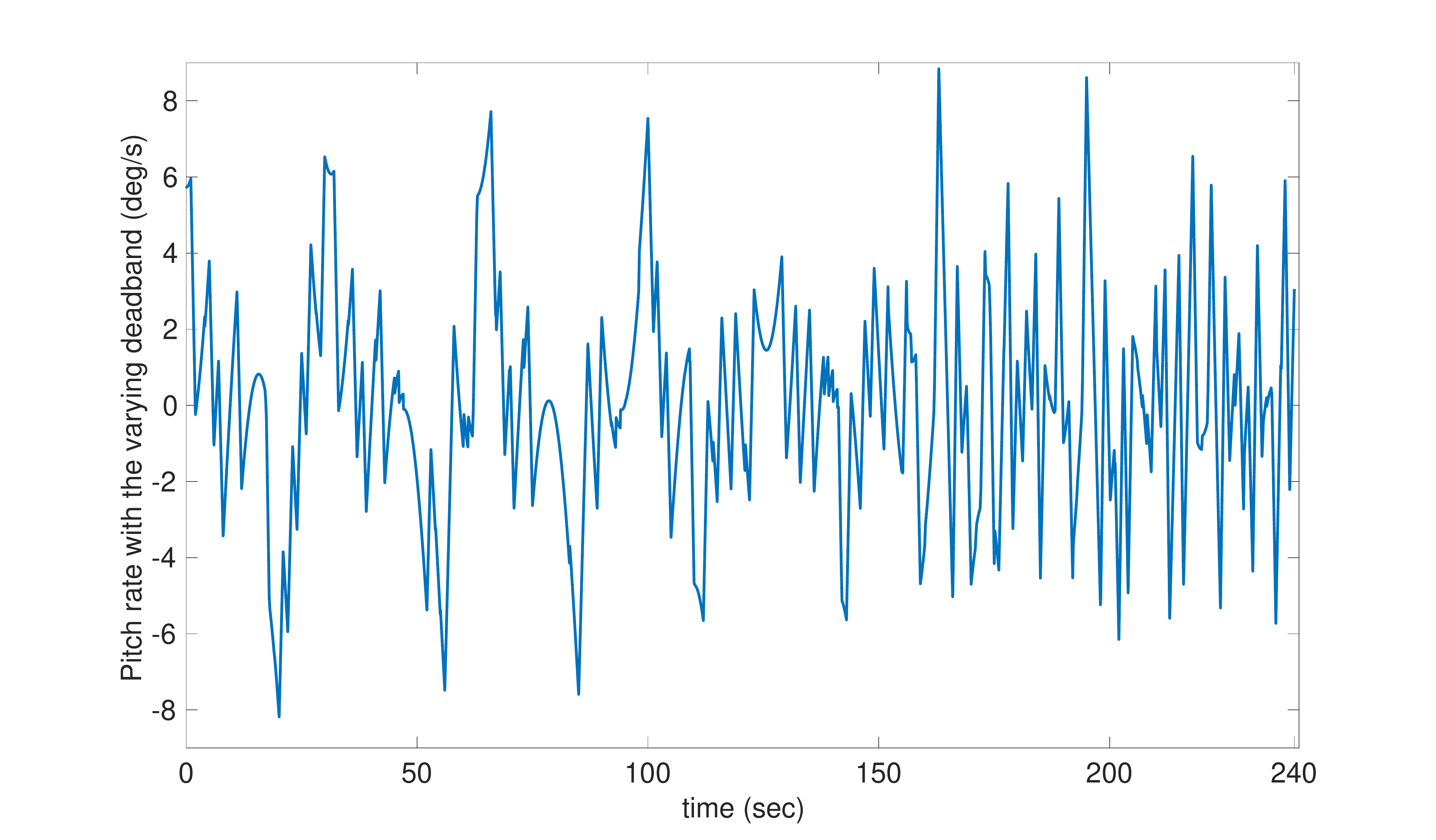}
    \caption{Pitch rate with the varying dead band}
    \label{fig:fig24}
\end{figure}

\begin{figure}[H]
    \centering
    \includegraphics[scale=0.25]{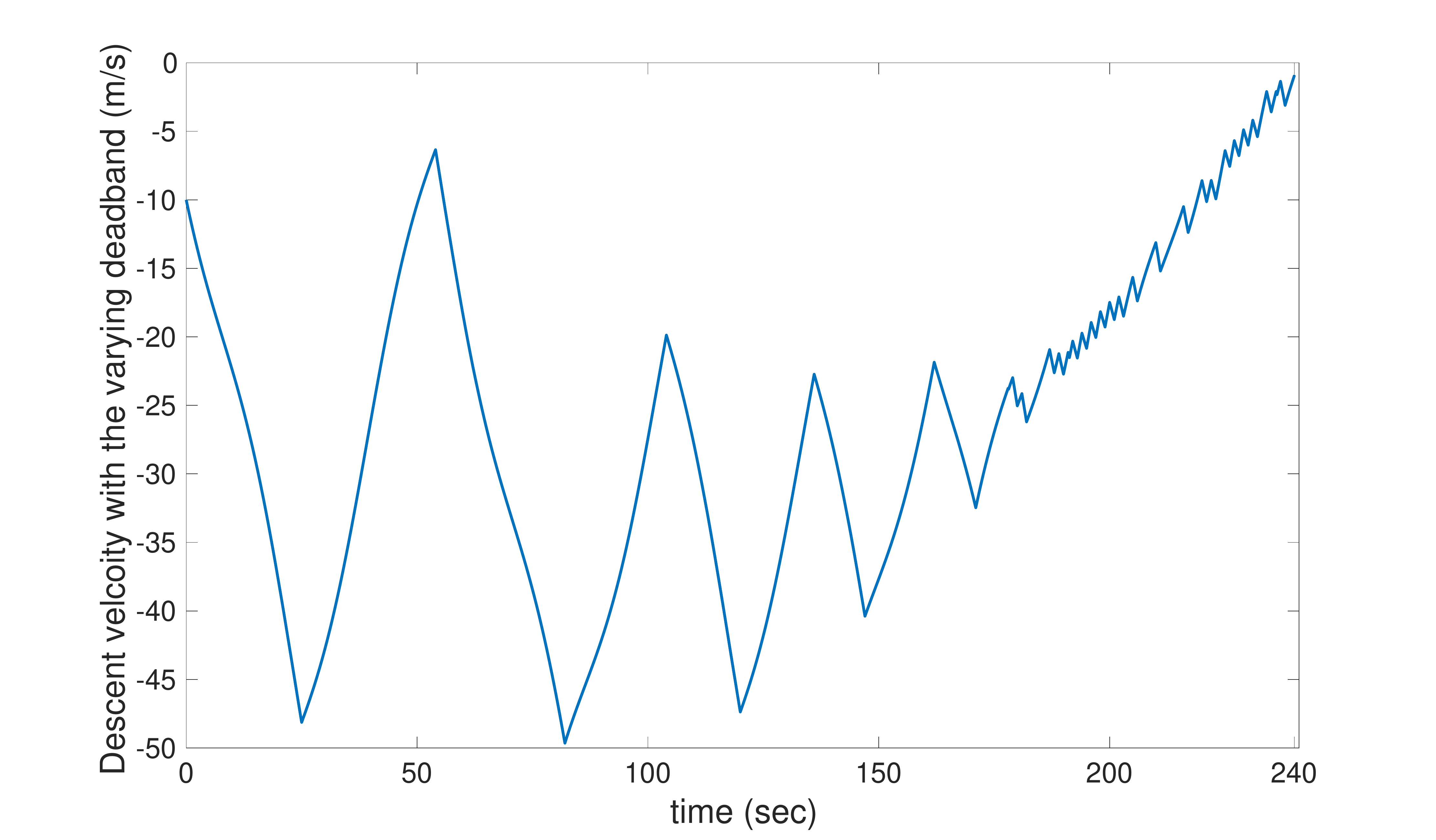}
    \caption{Descent velocity with the varying dead band}
    \label{fig:fig25}
\end{figure}

\begin{figure}[H]
    \centering
    \includegraphics[scale=0.25]{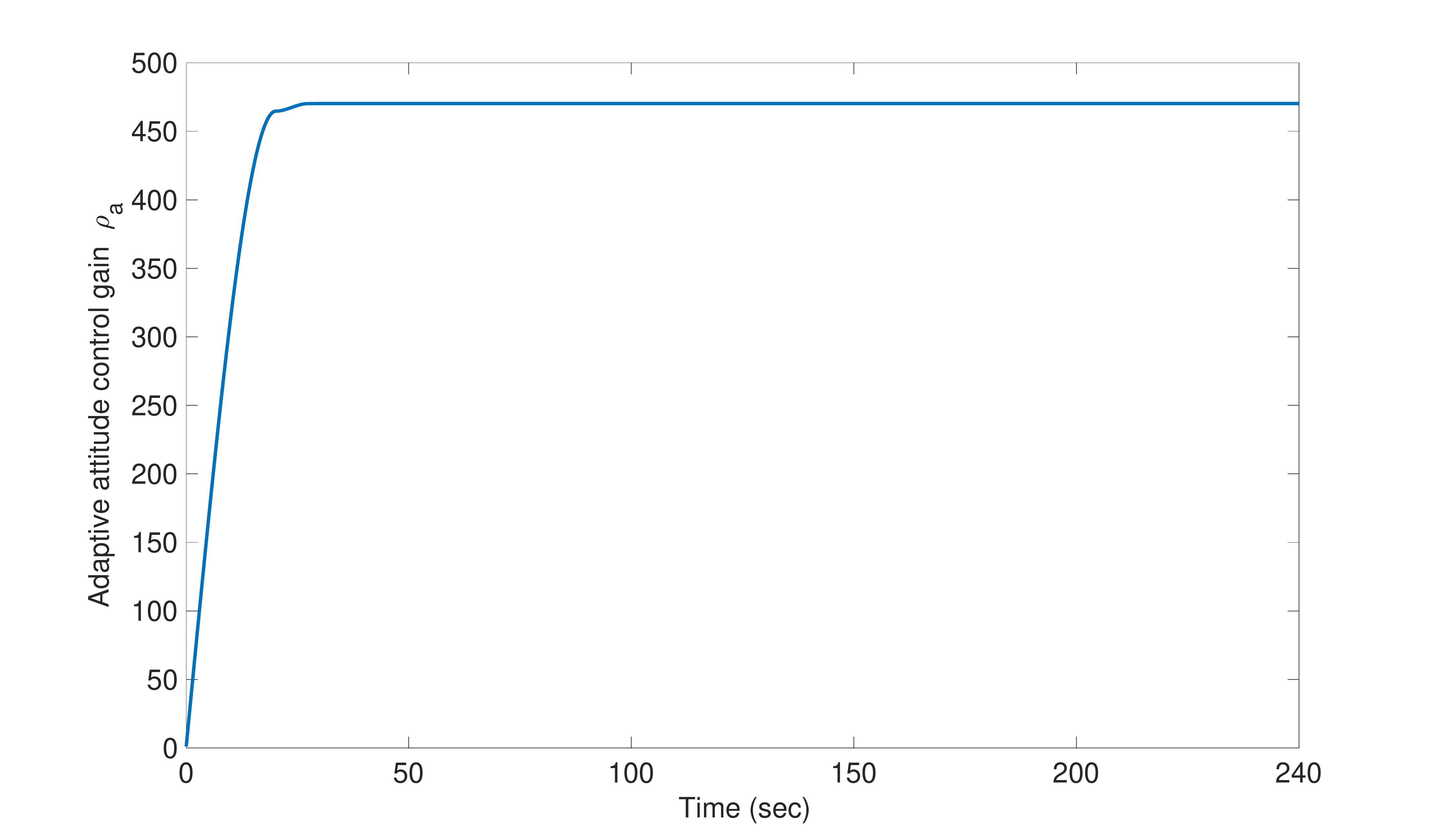}
    \caption{Adaptive attitude control gain \(\rho_{a}\) with the varying dead band}
    \label{fig:fig26}
\end{figure}

\begin{figure}[H]
    \centering
    \includegraphics[scale=0.25]{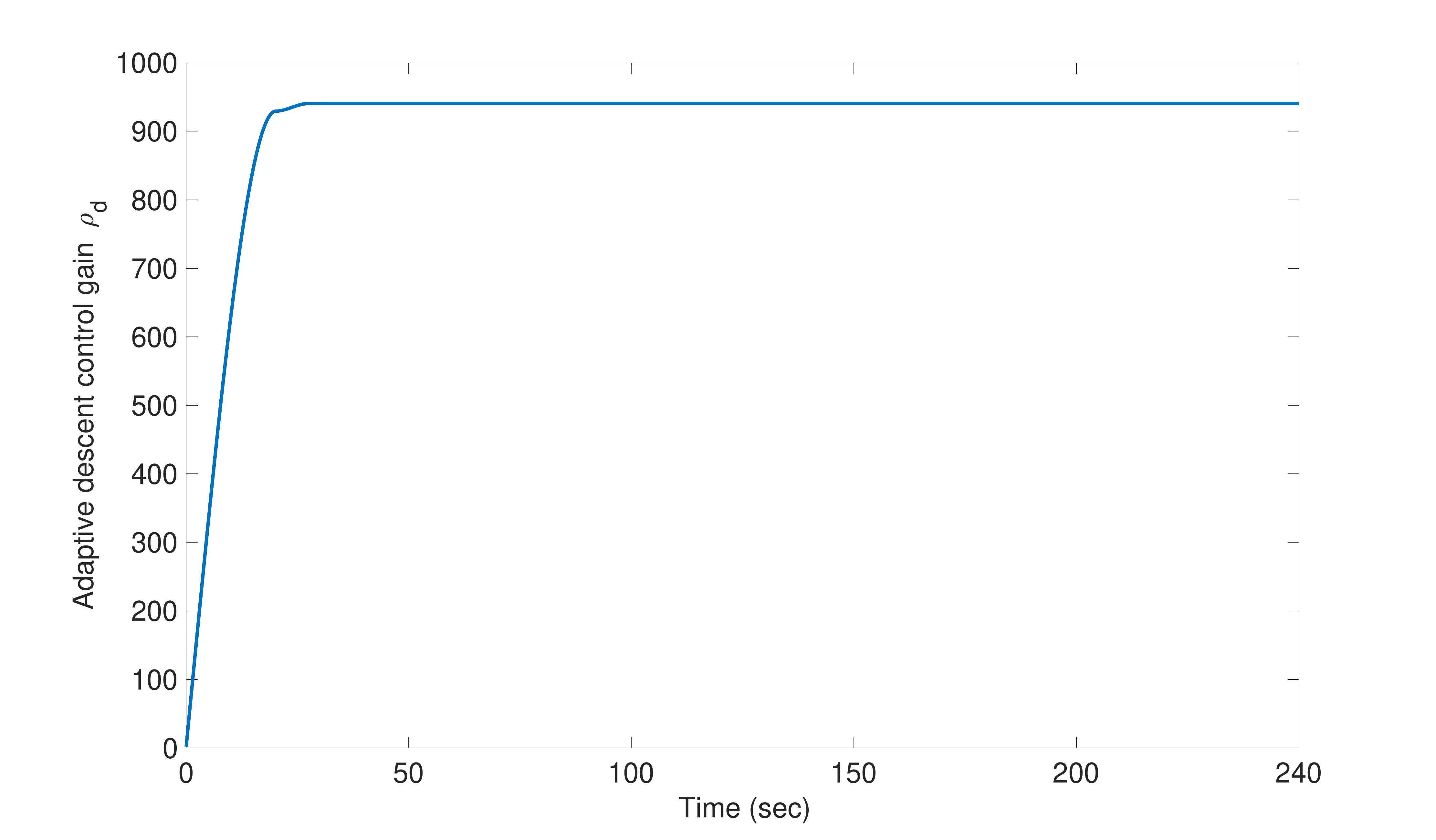}
    \caption{Adaptive descent control gain \(\rho_{d}\) with the varying dead band}
    \label{fig:fig27}
\end{figure}

The simulation results obtained using the \emph{SMC Aero} toolbox, of RPL in a soft lunar landing controlled by the adaptive 1-SMCs are presented in Figures~\ref{fig:fig18}-\ref{fig:fig27}.

\vspace{-0.7cm}

\vspace{-0.7cm}

\vspace{-0.7cm}

\emph{\textbf{Discussion:}}

\begin{itemize}
\item
  From Figures~\ref{fig:fig18} and \ref{fig:fig19}, one can see the pitch angle and altitude commands are followed accurately, while the soft lunar landing is provided. It is observed that, due to a varying dead band, the accuracy of tracking is improved by the landing time \(T = 240s\).
\item
  The plots of Figures~\ref{fig:fig20} and \ref{fig:fig21} show the sparse attitude and descent/altitude control functions with the increased switching
  frequency by the end of the landing due to decreasing the dead bands that aims to improve the accuracy of trajectories tracking; the duration of control pulses are not less than \(50\ ms\) as required.
\item
  From Figures~\ref{fig:fig22} and \ref{fig:fig23}, one can see that pitch sliding variable and descent sliding variable with varying dead band goes to small domains in a vicinity of zero by the time of landing.
\item
  The RPL pitch rate, \(\dot{\theta}\), and the decent velocity, \(\dot{x}\), presented in Figures~\ref{fig:fig24} and \ref{fig:fig25}, are with in the reasonable limits \(\left| \dot{\theta}(t) \right| \leq \ 1.14\ (deg/s)\),\(\ \left| \dot{x}(t) \right| \leq \ 0.5\ (m/s)\) by the landing time \(T = 240s\) while providing a soft landing with a vertical orientation of RPL.
\item
  The plots of the adaptive control gains \(\rho_{a}, \rho_{d}\) shown in Figures~\ref{fig:fig26} and \ref{fig:fig27}, demonstrate a quick gain adaptation (self-tuning) while constraining the sliding variables to the varying domains \( |\sigma_a| \leq \delta_{ia}, |\sigma_d| \leq \delta_{id}, \: i = 1,2,3 \) in accordance with Table~\ref{table:1}.
\end{itemize}

\subsection{Comparison with PID controllers and Adaptive Super Twisting controllers}

In this subsection we study the performances of

(a) PID controllers \cite{ref20} given by

\begin{equation}\label{eq:57}
\begin{aligned}
    u_{a PID} = 10e_{a}(t)+0.1\int_{0}^{t}e_{a}(t)d\tau+0.5\dot{e}_{a}(t) \\
    u_{d PID} = 30e_{d}(t)+0.2\int_{0}^{t}e_{d}(t)d\tau+0.1\dot{e}_{d}(t)
\end{aligned}
\end{equation}

and

\begin{figure}[H]
    \centering
    \includegraphics[scale=0.25]{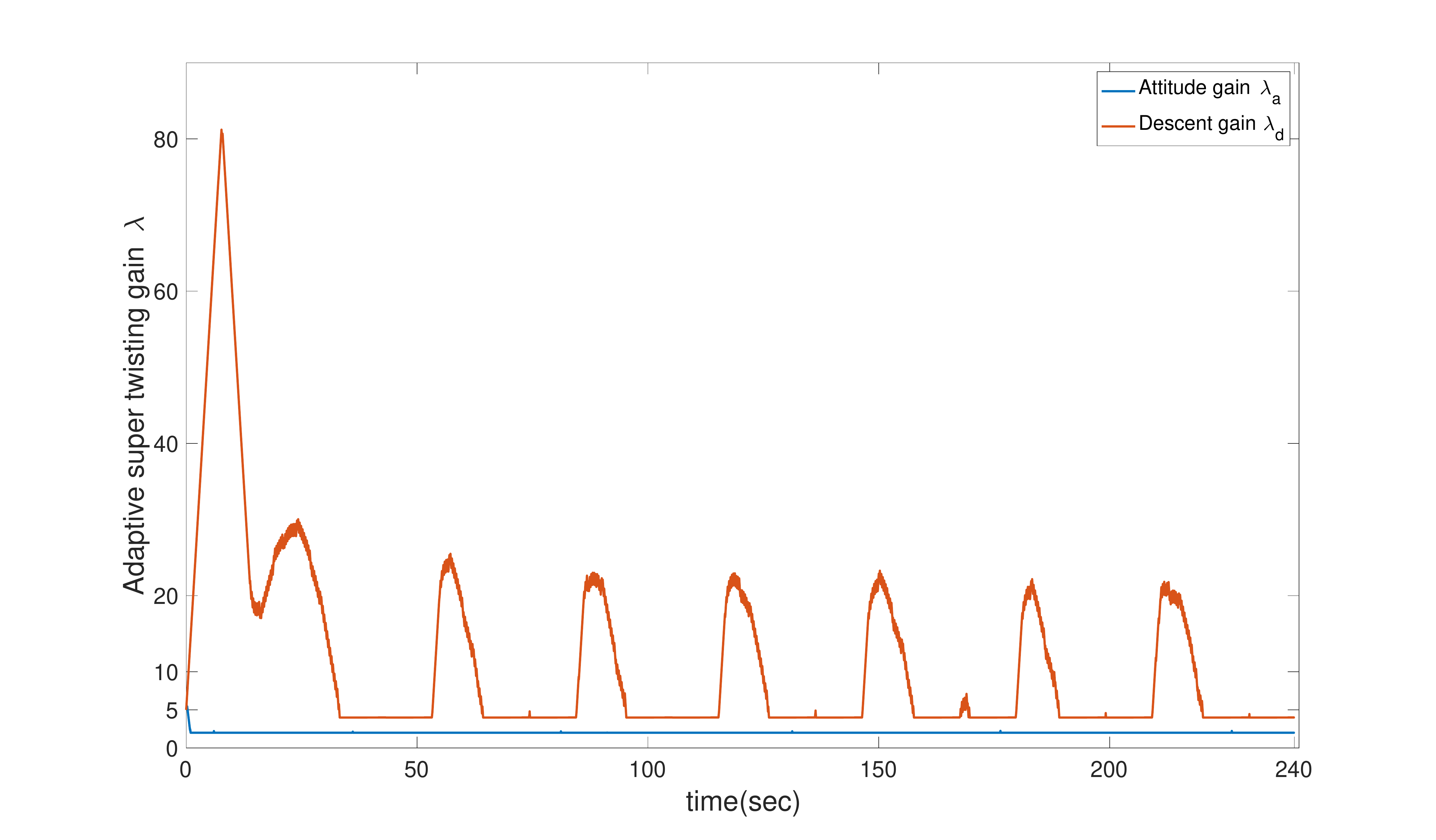}
    \caption{Adaptive Super Twisting gains \(\lambda_{a}\) and \(\lambda_{d}\) with the varying dead band}
    \label{fig:fig28}
\end{figure}

\begin{figure}[H]
    \centering
    \includegraphics[scale=0.25]{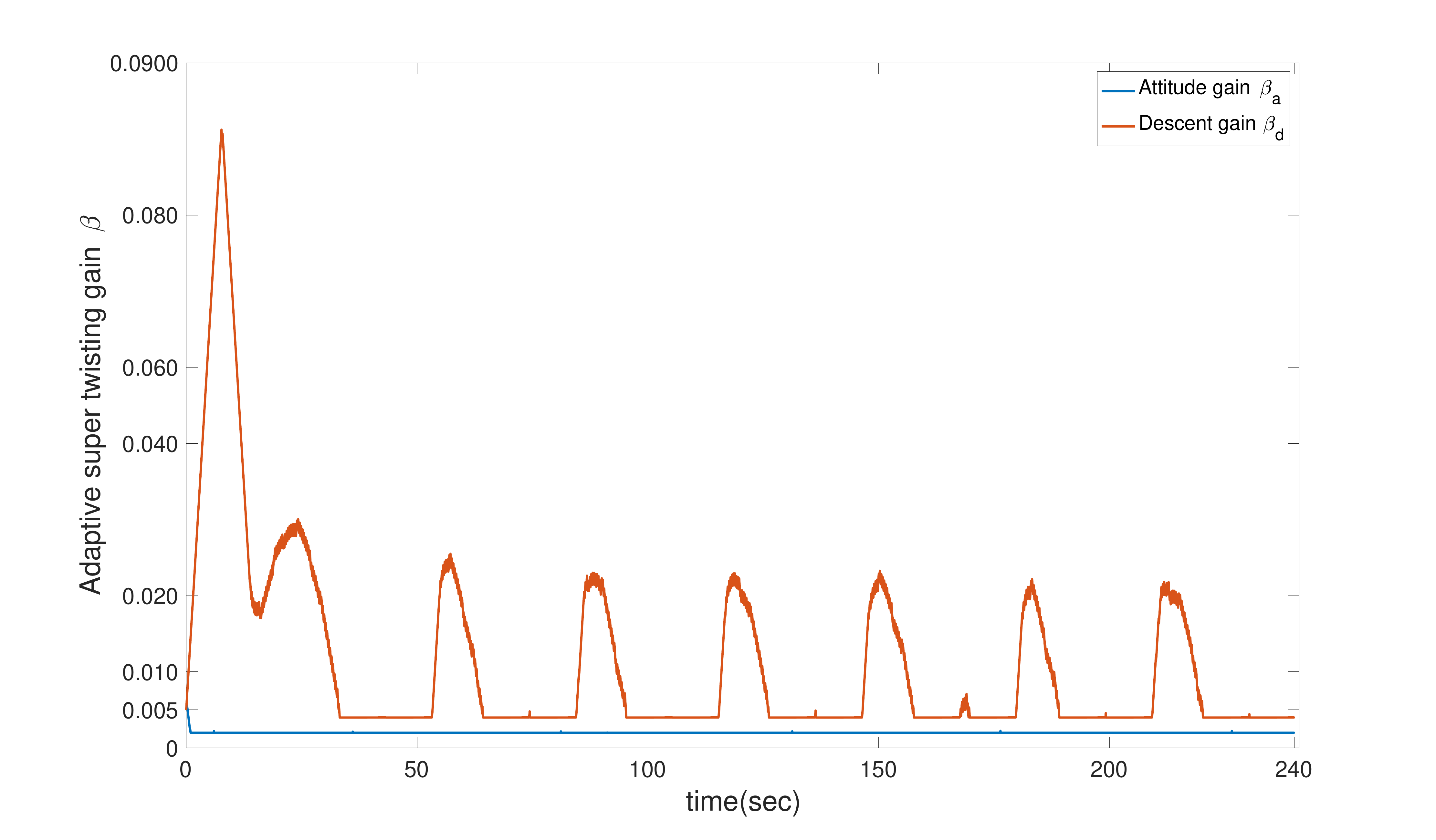}
    \caption{Adaptive Super Twisting gains \(\beta_{a}\) and \(\beta_{d}\) with the varying dead band}
    \label{fig:fig29}
\end{figure}

(b) Adaptive Super Twisting (STW) controllers \cite{ref7} in (\ref{eq:21}) and (\ref{eq:24}) as \(u_{\text{a\ STW}}\) and \(u_{\text{d\ STW}}\) designed using Super Twisting 2-SMC Adaptive Control block. Both controllers that drive the perturbed RPL in (\ref{eq:46}) in the lunar landing mode are implemented in the pulse width modulated format with S\&H blocks. The adaptive gains of the STW controllers are shown in Figure~\ref{fig:fig28} and Figure~\ref{fig:fig29}, where the possibility of the adaptive control gain \(\lambda\) and \(\beta\) reduction is demonstrated.

All three controllers demonstrated a reasonable performance in RPL soft lunar landing. The controllers' performance was compared using the following criteria:

(a) \emph{The Average Integral} \emph{Absolute} \emph{Attitude and Descent Errors} and over the time interval from zero until \(T = 240\ s\) that characterize the avarage accuracies:

\begin{equation}\label{eq:58}
    J_{e_{a}} = \frac{1}{T}\int_{0}^{T}|e_{a}(t)|d\tau, \enspace J_{e_{d}} = \frac{1}{T}\int_{0}^{T}|e_{d}(t)|d\tau 
\end{equation}

(b) \emph{The Average Integral} \emph{Absolute Controls} and over the time interval from zero until \(T = 240\ s\) that characterize average fuel spendings per second:

\begin{equation}\label{eq:59}
    J_{u_{a}} = \frac{1}{T}\int_{0}^{T}|u_{a}(t)|d\tau, \enspace J_{u_{d}} = \frac{1}{T}\int_{0}^{T}|u_{d}(t)|d\tau 
\end{equation}

The comparison results of PID, adaptive STW controllers and adaptive 1-SMC controllers in a pulse width modulated formats are presented in Table~\ref{table:2}.

\begin{table}[h!]
\centering
\caption{The comparison results of PID, adaptive Super Twisting controllers and adaptive 1-SMC controllers.}
\vspace{0.3cm}\begin{tabular}{c c c c c} 
 \hline
 Controllers & \(J_{e_{a}}\) & \(J_{e_{d}}\) & \(J_{u_{a}}\) & \(J_{u_{d}}\) \\ [0.5ex] 
 \hline
PID & 0.5855 & 336.6 & 5.304 & 10220\tabularnewline
STW & 0.0463 & 3.427 & 3.054 & 2552\tabularnewline
1-SMC & 0.1234 & 66.58 & 5.206 & 1513\tabularnewline
 \hline
\end{tabular} 
\label{table:2}
\end{table}

\subsubsection{Discussion}

Based on the results presented in Table~\ref{table:2} it can be observed that

\begin{itemize}
\item
  both adaptive 1-SMC and STW controllers outperformed the PID controller in either average fuel spending or average accuracy.
\item
  The adaptive STW controller outperforms the adaptive 1-SMC in the descent and the attitude average accuracy, and the attitude average fuel spending, while the adaptive 1-SMC shows better performance with respect to adaptive STW controller in in the average descent fuel spending.
\end{itemize}

\section{Second Case Study: Launch Vehicle Control and Simulation \cite{ref44}}
A purpose of this section is to demonstrate the capabilities of the developed \emph{SMC Aero} toolbox on the continuous 1-SMC design and simulation for a launch vehicle (Figure~\ref{fig:fig2}).

The designed continuous 1-SMC is compared to PD controller presented in the work \cite{ref33}.

\subsection{LV mathematical modeling \cite{ref32,ref33}}

A simplified mathematical pitch plane attitude dynamics of an LV in the ascent mode in the presence of the bounded perturbations are given by \cite{ref33}:

\begin{equation}\label{eq:60}
    I\ddot{\theta} = C_{N\alpha}\bar{q}Sl_{a}\theta - \frac{C_{N\alpha}\bar{q}Sl_{a}}{V}\dot{z} - l_{g}R\beta -(l_{g}S_{n}+I_{n})\ddot{\beta}
\end{equation}

\begin{equation}\label{eq:61}
    M\ddot{z} - M\bar{g}\theta = C_{N\alpha}\bar{q}S\theta - \frac{C_{N\alpha}\bar{q}S}{V}\dot{z} + R\beta+S_{n}\ddot{\beta}
\end{equation}

where, \(\theta(deg)\) is a pitch perturbation attitude, \(\dot{\theta}\ (deg/s)\) pitch rate, \(\beta\left( \deg \right)\ \) is the gimbal angle and \(\dot{z}\ (deg/s)\ \)is the lofting velocity.

The parameters that contribute to the LV dynamics are presented as \(I = 1.355 \times 10^{8}\ (kg - m^{2}\)) is the pitch axis principal moment of inertia;\(\text{\ C}_{N\alpha} = 8.0\) is the pitching normal force coefficient; \(\bar{q} = 4880\ (kg/m^{2})\) is a dynamic pressure; \(S = 9.2\left( m^{2} \right)\) is an aerodynamic reference area; \(l_{a} = 15.24(m)\) is center of pressure location; \(S_{n} = 135.58\ (kg - m^{2})\) is the first moment of inertia of engine about gimbal; \(V = 457.2(m/sec)\) is the Earth-relative velocity magnitude; \(l_{g} = 22.86(m)\) is the gimbal location; \(R = 0.4535 \times 10^{6}\ \left( \text{kg} \right)\) is the vectored thrust; \(\bar{g} = 18.29\ (m/\sec^{2})\) is the mean axial acceleration; \(M = 21.89 \times 10^{4}\ (kg)\) is the vehicle mass and \(I_{n} = 1355.8\ (kg - m^{2})\) is the second moment of inertia.

Note that in equations (\ref{eq:60}) and (\ref{eq:61}) the LV translational dynamics are ignored and only the short-period pitch dynamics are considered for this simplified study, whose goal is to demonstrate capabilities of the \emph{SMC Aero} toolbox to aerospace vehicle sliding mode control design and simulation.

The input-output pitch plane dynamics of an LV in (\ref{eq:60}) and (\ref{eq:61}) is presented in a transfer function format, while taking into account actuator and flexible mode dynamics;

(a) The \emph{rigid body dynamics}, including the effects of inertial coupling which is represented by a second-order transfer function is given by

\begin{equation}\label{eq:62}
    G_{1}(s) = \frac{\theta(s)}{\beta(s)} = \frac{-(l_{g}S_{n}+I_{n})s^{2}-l_{g}R}{Is^{2}-C_{N\alpha}\bar{q}Sl_{a}}
\end{equation}

(b) The \emph{simple actuator dynamics,} which is represented by a combination of a second-order engine pendulum dynamics and the first-order response of a hydraulic servo actuator, is given by

\begin{equation}\label{eq:63}
    G_{a}(s) = \frac{\beta(s)}{u_{\theta}(s)} = (\frac{1}{\tau_{a}s+1})(\frac{{\omega_{e}}^{2}}{s^{2}+2\zeta_{e}\omega_{e}s+{\omega_{e}}^{2}})
\end{equation}

where, \(\omega_{e} = 12\ (rad/sec)\ \)is the natural frequency of the engine with the servo loop closed; \(\zeta_{e} = 0.5\) is the damping ratio of the engine with servo loop closed and \(\tau_{a} = 0.2\ (sec)\) is the actuator time constant; \(u_{\theta}\) is a control function.

(c) The \emph{bending (flexible mode) dynamics}, which is represented by a second order transfer function is given by

\begin{equation}\label{eq:64}
    G_{b}(s) = \frac{\theta(s)}{\beta(s)} = \psi_{s}\bigg\{ \frac{(S_{n}\phi_{g}-I_{n}\psi_{g})s^{2}+R\phi_{g}}{s^{2}+2\zeta_{b}\omega_{b}s+{\omega_{b}}^{2}} \bigg \}
\end{equation}

where, \(\omega_{b} = 12.5\ (rad/sec)\) is the natural frequency of the flexible mode and \(\zeta_{b} = 0.005\) is the damping ratio of the flexible mode; \(\phi_{g} = - 0.0005\) is the mass normalized mode shape at the gimbal; \(\psi_{g} = 0.0001\) is the slope at the gimbal and \(\psi_{s} = 0.0001\) is the mode slope at the sensor location.

Based on equations (\ref{eq:62}) - (\ref{eq:64}) the open-loop transfer function of an LV defined by a transfer function is given in \cite{ref33}

\begin{equation}\label{eq:65}
    G_{OL}(s) = \frac{\theta(s)}{u_{\theta}(s)} = G_{a}(G_{1}+G_{b})
\end{equation}

\begin{figure}[H]
    \centering
    \includegraphics[scale=0.2]{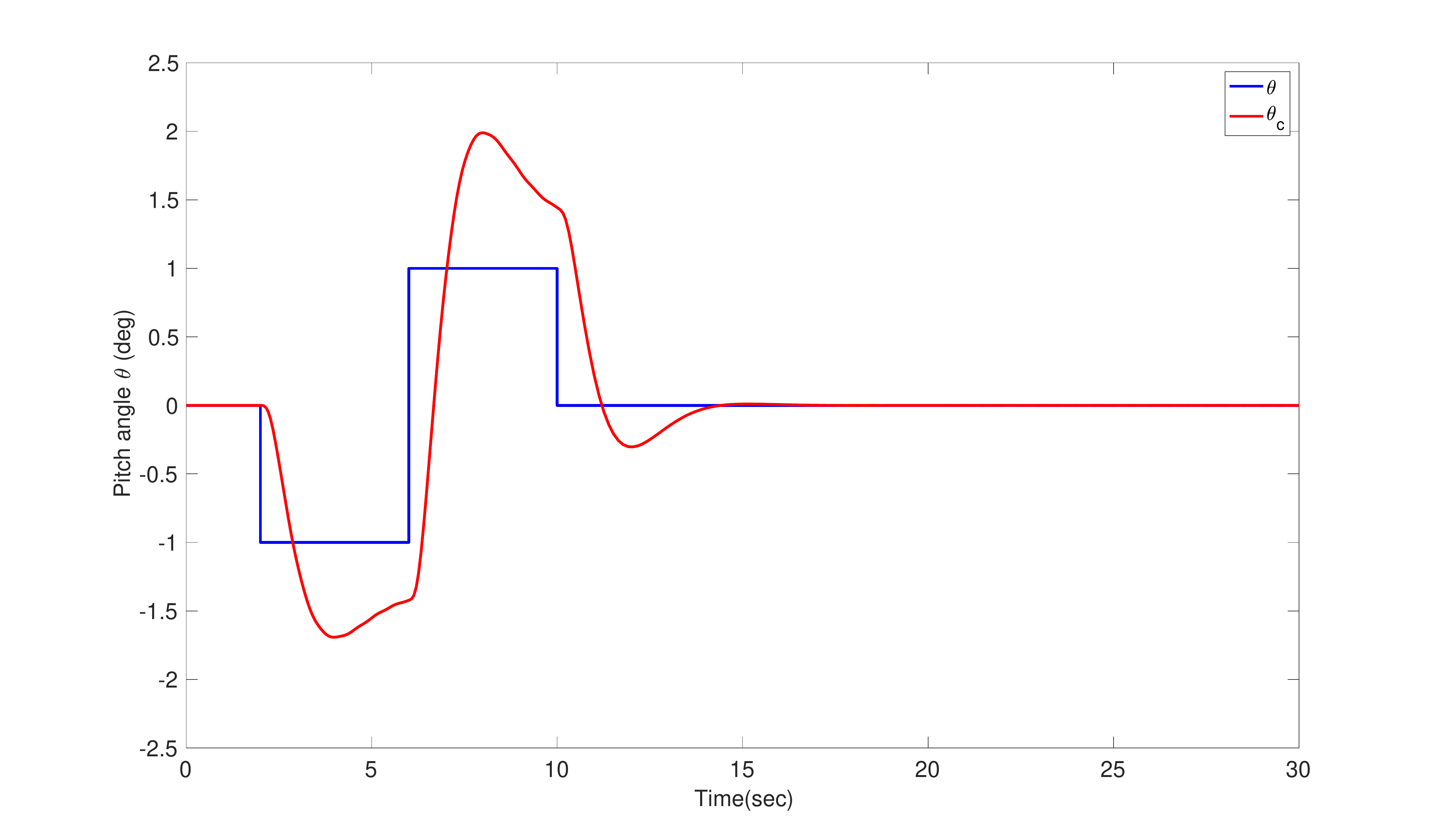}
    \caption{Transient response of the unperturbed closed loop system with the PD controller (\ref{eq:66}), (\ref{eq:67})}
    \label{fig:fig30}
\end{figure}

The cascade PD-controller with a transfer function

\begin{equation}\label{eq:66}
    H(s) = \frac{u_{\theta}(s)}{e_{\theta}(s)} = k_{p}+k_{d}s
\end{equation}

where \( e_{\theta} = \theta_{c}-\theta\) \enspace (\(\theta_{c}\) is the command attitude trajectory) and

\begin{equation}\label{eq:67}
\begin{aligned}
    k_{p} = \frac{I{\omega_{n}}^{2}+C_{N\alpha}Sl_{a}\bar{q}}{Rl_{g}} = 1.87 \\
    k_{d} = \frac{2\zeta\omega_{n}I}{Rl_{g}} = 2.13, \omega_{n}= 1, \zeta = 0.7 
\end{aligned}
\end{equation}

was proposed in \cite{ref33} to control the pitch angle \(\theta\) in unperturbed system with the open loop transfer function (\ref{eq:65}). The transient response is presented in Figure~\ref{fig:fig30} with

\begin{equation}\label{eq:68}
    \theta_{c}(t) = -1.2 \cdot 1(t-2) + 2.3 \cdot 1(t-6) - 1.2 \cdot 1(t-10)
\end{equation}

The linear controller bandwidth is limited by the design method which ensures classical stability margins. The transient response is reasonably good, but the accuracy of \(\theta_{c}\left( t \right)\) following can be improved, especially in a perturbed case.

\subsection{LV SMC Design using SMC Aero toolbox}

The two SMC controllers are designed in terms of \(u_{\theta}\):

(a) the \emph{continuous quasi SMC} that drives the tracking error \(e_{\theta} = \ \theta_{c} - \theta\) to the domain \(\left| e_{\theta} \right| \leq \ \varepsilon_{\theta}\), \(\varepsilon_{\theta} > 0\ \) by the operation time \emph{T};

(b) the \emph{2-SMC STW} controller: 

The performances of the designed SMC controllers are compared with the PD controller proposed in \cite{ref33}. The continuous quasi SMC is designed using the \emph{SMC Aero} toolbox in the following steps:

\emph{Step 1:} Practical relative degree identification
The PRD ID block is used to identify the scalar practical degree \(r = \lbrack 2\rbrack\). The corresponding MATLAB editor code of a performance analyzer and PRD ID block with embedded mask feature are shown in Figures \ref{fig:fig4} and \ref{fig:fig5}.

\emph{Step 2:} Sliding variable design: Given \(m = 1\) and the desired settling time \(t_{s} = \ 8\ \left( \sec \right)\) in the attitude sliding mode, the sliding variable \(\sigma_{\theta}\) is designed in a form (\ref{eq:9}) using the block Sliding Variable Design, whose structure is presented in Figure~\ref{fig:fig8}. As a result, the following attitude (pitch angle) sliding variable is obtained:

\begin{equation}\label{eq:69}
    \sigma_{\theta} = \dot{e}_{\theta}+1.75e_{\theta}+1.56\int e_{\theta}dt 
\end{equation}

\emph{Step 3:} Continuous quasi SMC design and the implementation

The SMC control function \(u_{\theta}\) is introduced in accordance with eq. (\ref{eq:13}). The continuous (quasi) 1-SMC is coded in a continuous sigmoid format, where, \(\text{sign}\left( \sigma_{\theta} \right)\) terms in eq. (\ref{eq:13}) is replaced by \(\frac{\sigma_{\theta}}{\left| \sigma_{\theta} \right| + \varepsilon_{\theta}}\ \) with \(\varepsilon_{\theta} > 0\) being a small number. The continuous (quasi) 1-SMC controller \(u_{\theta}\) that drives \(\sigma_{\theta}\) to the domain \(\left| e_{\theta} \right| \leq \ \varepsilon_{\theta}\), \(\varepsilon_{\theta} > 0\ \)in finite time in the presence of the bounded disturbance \(|f_{\theta}(t)| \leq \bar{L}, \bar{L}>0\) computed as

\begin{equation}\label{eq:70}
    u_{\theta} = -(\rho_{\theta}+\bar{L})\frac{\sigma_{\theta}}{|\sigma_{\theta}|+\epsilon_{\theta}}
\end{equation}

Note that the fixed gain \(\rho_{\theta} > 0\) that guarantees a desired convergence time

\begin{equation}\label{eq:71}
    t_{r\theta} \leq \frac{|\sigma_{\theta}(0)|}{\rho_{\theta}}
\end{equation}

and the Lipschitz constant \(\bar{L}\) is supposed to be tuned.

\begin{figure}[H]
    \centering
    \includegraphics[scale=0.8]{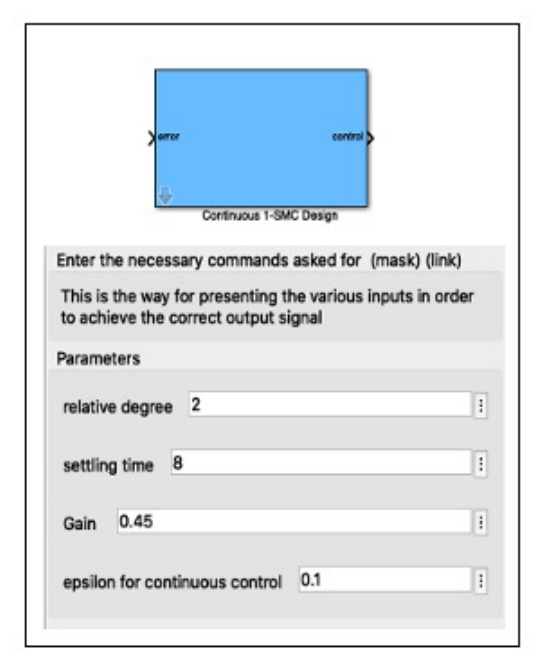}
    \caption{Continuous 1-SMC Design block with the mask overview}
    \label{fig:fig31}
\end{figure}

The continuous (quasi) 1-SMC is designed using SISO 1-SMC Fixed Gain block with \(m = 1\), whose structure and the mask overview are presented in Figure~\ref{fig:fig31}

Note that the adaptive STW controller \(u_{\theta STW}\) is designed in accordance with equations (\ref{eq:22}) and (\ref{eq:24}) and using the Super Twisting 2-SMC
Adaptive Control block.

\subsection{Simulation Setup:}\label{simulation-setup-1}

\begin{itemize}
\item
  The commanded \(\theta_{c}\left( t \right)\) trajectory is given in equation (\ref{eq:68}),
\item
  The controlled operation time is given as \(T = 30s\),
\end{itemize}

\begin{itemize}
\item
  The Parameters of the continuous 1-SMC controller are taken as:

  fixed control gain \(\rho_{\theta} = 0.45\), small positive number \(\varepsilon_{\theta}\) = 0.1.
\end{itemize}

\begin{itemize}
\item The perturbation
\begin{equation}\label{eq:72}
    f_{\theta}(t) = 0.2+0.1sin(0.1t) (rad)
\end{equation}

\end{itemize}

\begin{quote}
that is added to the input of the actuator (\ref{eq:65}) additively with the control function \(u_{\theta}\) is due to the thrust vector misalignment and is used to demonstrate the robustness of continuous (quasi) 1-SMC controller.
\end{quote}

\begin{itemize}
\item
  Euler integration method with a fixed-step size \(10^{-5}\text{\ s}\) is used for the system simulation.
\item
  The initial conditions are in the origin.
\end{itemize}

\subsection{Simulation Diagram}

\begin{figure}[H]
    \centering
    \includegraphics[scale=1]{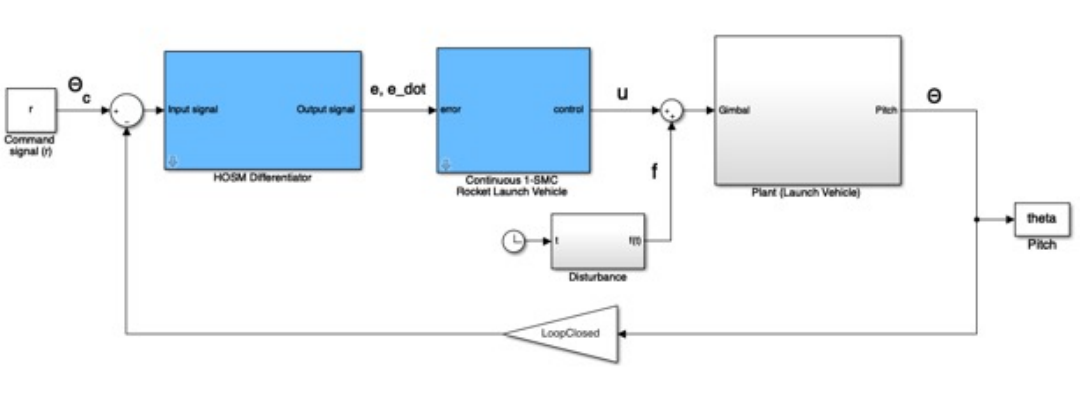}
    \caption{Simulation Diagram of a LV controlled by continuous 1-SMC controller}
    \label{fig:fig32}
\end{figure}

A simulation diagram of a LV controlled by a continuous 1-SMC controller in equation (\ref{eq:70}) is presented in the Figure~\ref{fig:fig32}.

\subsection{Simulation Results and Discussions:
}\label{simulation-results-and-discussions}

\begin{figure}[H]
    \centering
    \includegraphics[scale=0.25]{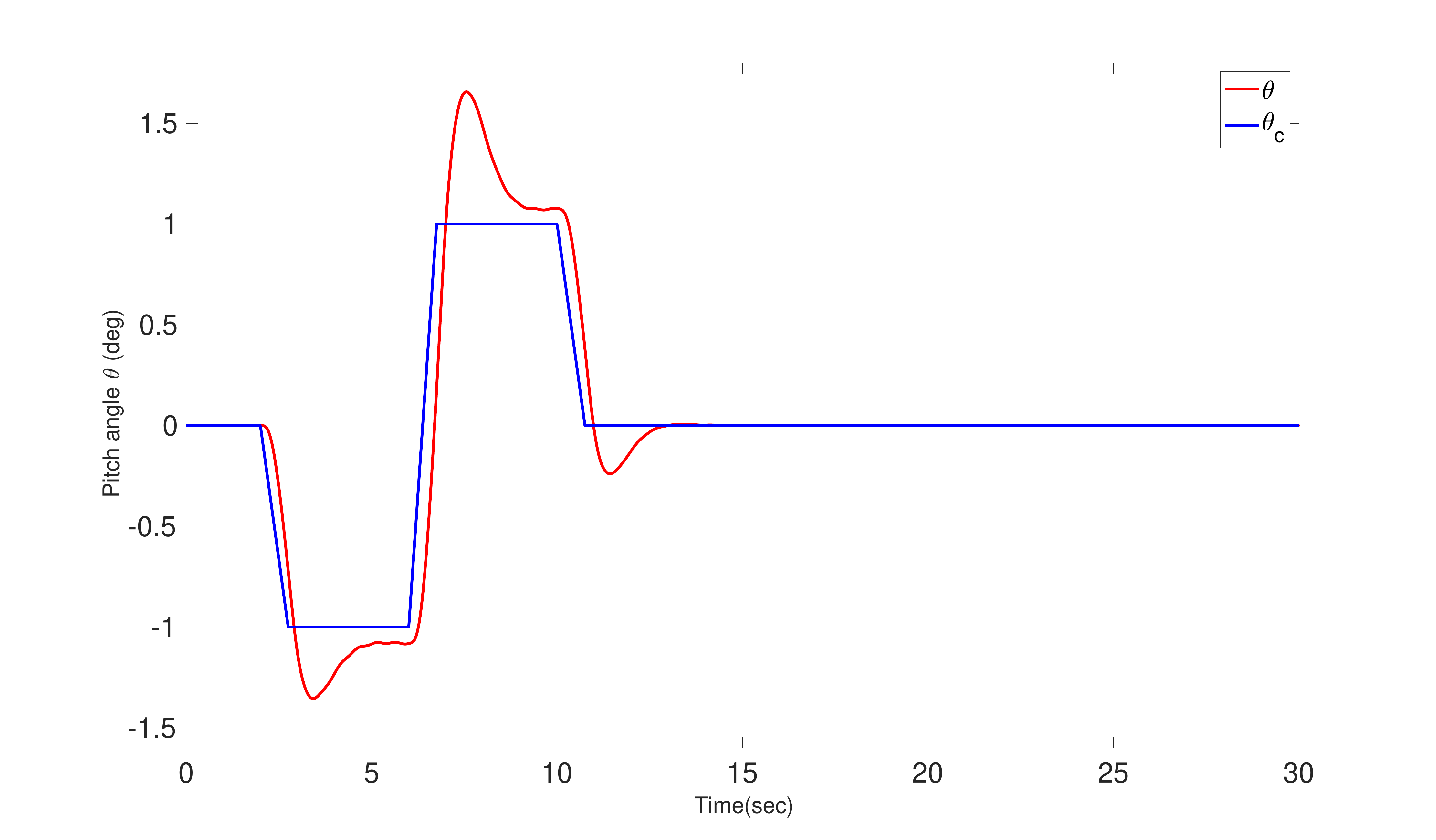}
    \caption{Pitch angle tracking via continuous 1-SMC controller}
    \label{fig:fig33}
\end{figure}

\begin{figure}[H]
    \centering
    \includegraphics[scale=0.25]{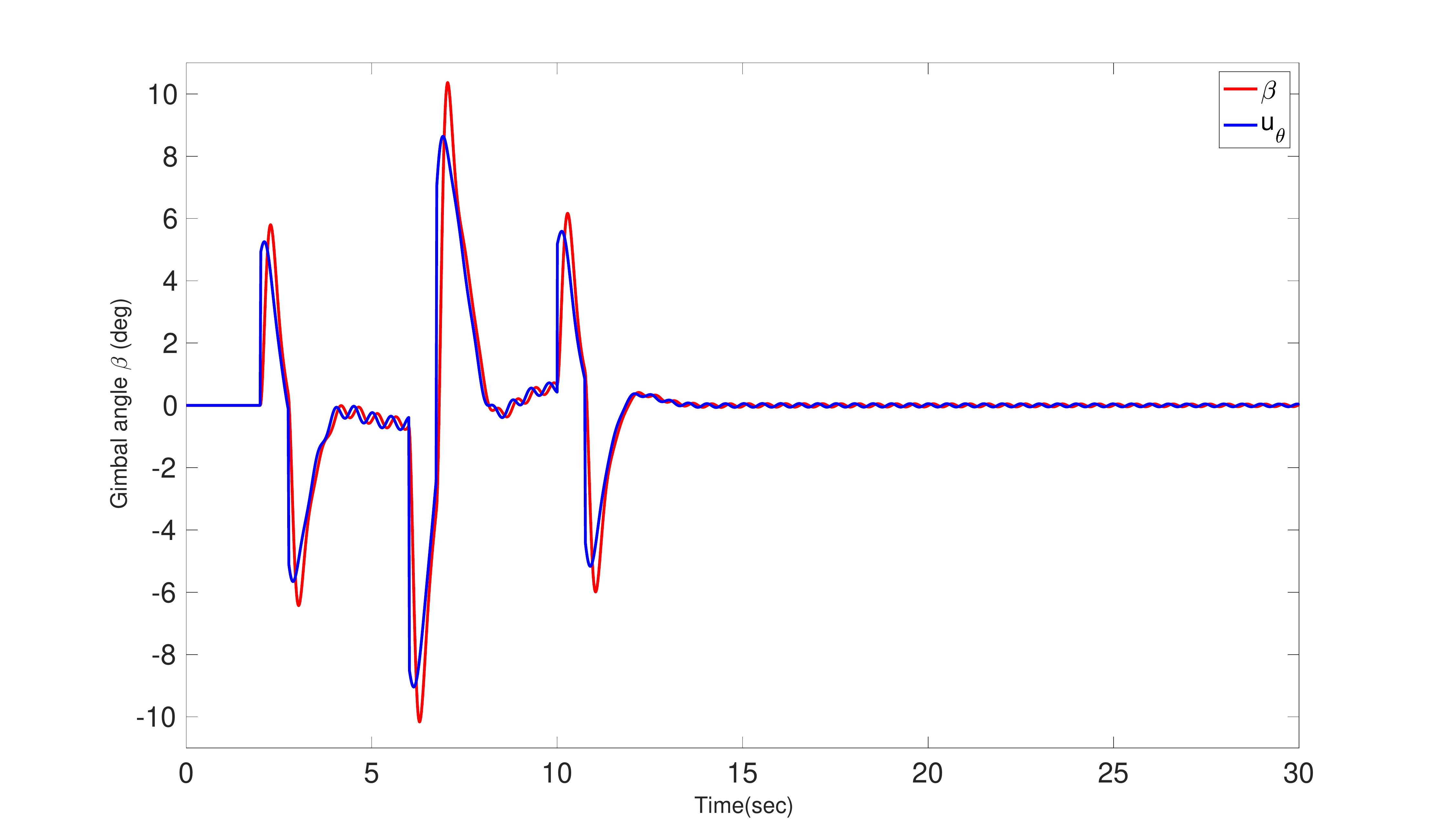}
    \caption{The control function \(u_{\theta}\) and the gimbal angle \(\beta(t)\)}
    \label{fig:fig34}
\end{figure}

The simulation results of LV in an ascent mode controlled by the continuous 1-SMC controller and obtained using the \emph{SMC Aero} toolbox are presented in Figures~\ref{fig:fig33} and \ref{fig:fig34}.

\emph{\textbf{Discussion:}}
The \emph{SMC Aero} toolbox demonstrates the strong efficacy in the LV continuous 1-SMC controller design that includes practical relative degree identification, sliding variable and control design, and the system simulation. From (Figure~\ref{fig:fig33}) and (Figure~\ref{fig:fig34}), one can see the excellent \(\theta\) tracking performance with the gimbal angle within a reasonable range. The designed continuous 1-SMC controller would serve as an autopilot in a rate-command mode with a synchronized attitude reference to maintain alignment with a zero angle of attack guidance profile until atmospheric exit.

\subsection{Comparison of continuous 1-SMC with PD and STW}

\begin{figure}[H]
    \centering
    \includegraphics[scale=0.2]{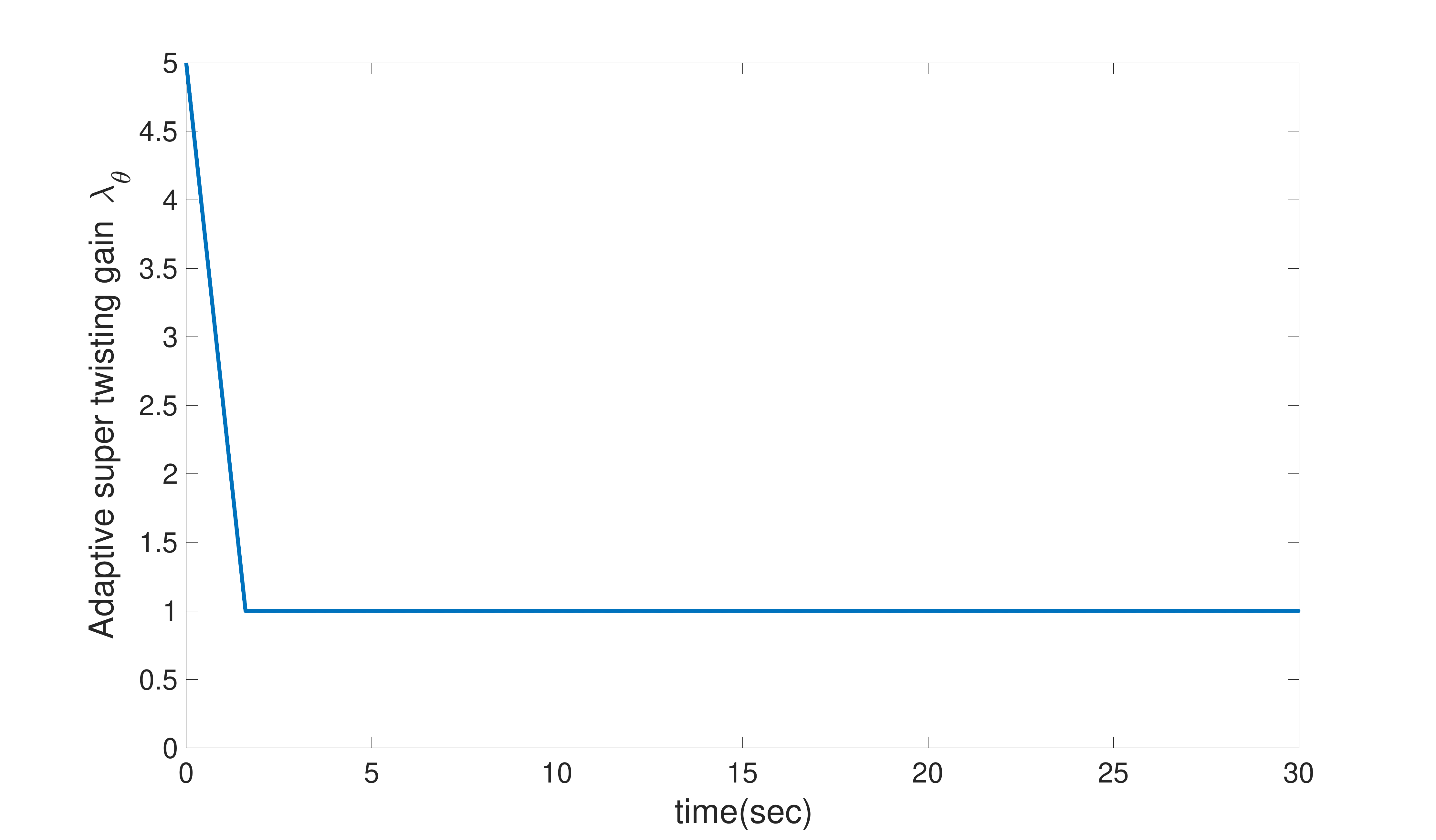}
    \caption{Adaptive Super Twisting attitude gain \(\lambda_{\theta}\)}
    \label{fig:fig35}
\end{figure}

\begin{figure}[H]
    \centering
    \includegraphics[scale=0.2]{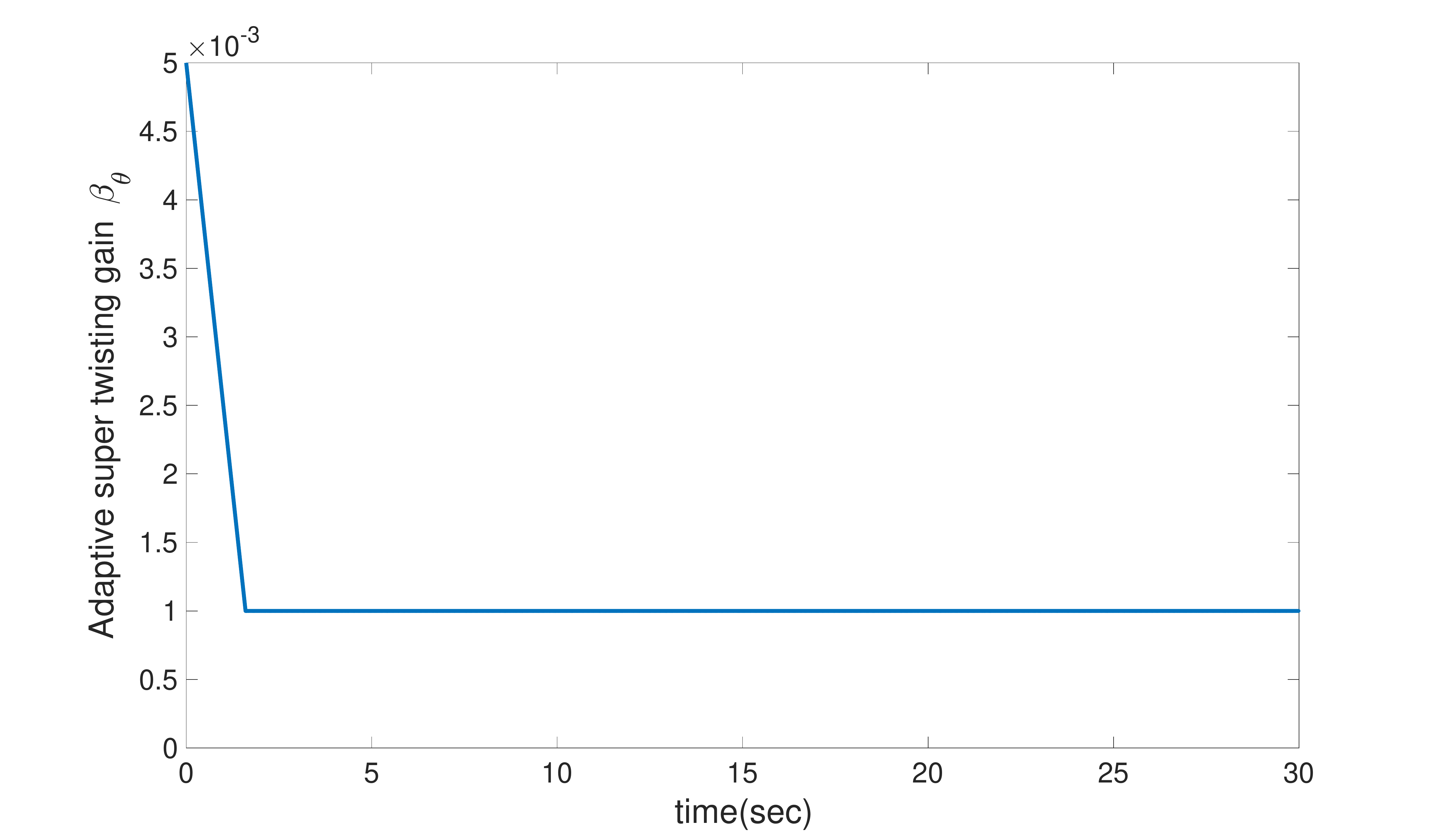}
    \caption{Adaptive Super Twisting attitude gain \(\beta_{\theta}\)}
    \label{fig:fig36}
\end{figure}

The performances of an adaptive STW controller \(u_{\theta STW}\) is comparable to continuous 1-SMC. The adaptive gains \(\lambda_{\theta}\) and \(\beta_{\theta}\) shown in Figures~\ref{fig:fig35} and \ref{fig:fig36} demonstrate self-tuning of the STW controller by dropping the gain values to a reasonable low-level preventing gain overestimation.

The performance of LV during the ascent mode controlled by PD controller in (\ref{eq:66}) and (\ref{eq:67}) is assessed (see Figure~\ref{fig:fig30}).

\subsection{Performance comparison: Results and Discussions}

The comparison criteria of PD controller in (\ref{eq:66}), (\ref{eq:67}) \cite{ref33}, continuous 1-SMC, and adaptive STW controller were selected as:

\begin{quote}
(a) \emph{The Average Integral Absolute} \emph{Attitude Error} \(e_{\theta}\) over the time interval from zero until \(T = 30\ s\) that characterizes the average accuracy:
\end{quote}

\begin{equation}\label{eq:73}
    J_{e_{\theta}} = \frac{1}{T}\int_{0}^{T}|e_{\theta}(t)|d\tau
\end{equation}

\begin{quote}
(b) \emph{The Average Integral Absolute Control} \(u_{\theta}(t)\) over the time interval from zero until \(T = 30\ s\) that characterizes the "fuel" usage (battery or hydraulic fluid usage).
\end{quote}

\begin{equation}\label{eq:74}
    J_{u_{\theta}} = \frac{1}{T}\int_{0}^{T}|u_{\theta}(t)|d\tau
\end{equation}

The comparison results of PD, adaptive Super Twisting controller and continuous 1-SMC controller are presented in Table~\ref{table:3}.

\begin{table}[h!]
\centering
\caption{The comparison results of PD, adaptive Super Twisting controller and continuous 1-SMC controller.}
\vspace{0.3cm}\begin{tabular}{c c c c} 
 \hline
 Controllers & \(J_{e_{\theta}}\) & \(J_{u_{\theta}}\) \\ [0.5ex] 
 \hline
PD & 0.04885 & 0.1119\tabularnewline
STW & 0.001371 & 0.09041\tabularnewline
1-SMC & 0.001602 & 0.01462\tabularnewline
 \hline
\end{tabular} 
\label{table:3}
\end{table}

\subsubsection{Discussion}

Based on the results presented in Table~\ref{table:3} it can be observed that

\begin{itemize}
\item
  both adaptive 1-SMC and STW controllers outperformed the PD controller in either average fuel spending or average accuracy.
\item
  The adaptive STW controller outperforms the adaptive 1-SMC in the average accuracy, while the adaptive 1-SMC shows better performance
  with respect to adaptive STW controller in the average fuel spending.
\end{itemize}

\section{Conclusion}
Sliding mode control and observation techniques are widely used in aerospace applications, including aircraft, UAVs, launch vehicles, missile interceptors, and hypersonic missiles. In this work a sliding mode controller design and simulation software toolbox that aims to support the aerospace vehicle applications was created for the first time in the MATLAB/Simulink platform. 

The developed \emph{SMC Aero} toolbox contain libraries that (a) employ the relative degree approach that allows supporting the robust to the bounded disturbances sliding mode controller design for aerospace vehicles treating the vehicle input-output dynamics as a "black box"; (b) are capable to support the design of the conventional and higher order sliding mode controllers and differentiators given required specification of the aerospace vehicle closed loop dynamics; (c) support simulations of the designed sliding mode control system; (d) are independent from each other and are rather task oriented. The proposed toolbox architecture/structure contains the libraries namely Practical Relative Degree ID, 1-SMC, 2-SMC, HOSM Control (HOSMC), Adaptive 1-SMC/2-SMC/HOSMC, HOSM Differentiators, Lunar Lander (RPL) 1-SMC and 2-SMC (Super Twisting Control), Launch Vehicle (LV) 1-SMC and 2-SMC (Super Twisting Control), and Applications. 

All implemented algorithms are based on the relative degree approach that requires minimal knowledge about the aerospace vehicle dynamics. The developed toolbox is verified on two cases study: controlling perturbed Resource Prospector Lander (RPL) in the powered descent mode on the Moon and controlling attitude of the perturbed Launch Vehicle (LV) in the ascent mode, whose simplified mathematical models are provided by NASA/MSFC. The performance comparison results of PID/PD controllers versus adaptive Super Twisting and fixed gain and adaptive 1-SMC controllers (designed using the created toolbox) were studied, analyzed and compared for both RPL and LV cases. The architecture of the \emph{SMC Aero} toolbox that facilitates the aerospace controller designs and simulations can be extended and expanded aiming at the improvement of the toolbox characteristics.

\section*{Acknowledgment}
This work is supported by NASA Marshall Space Flight Center's (MSFC) Control System Design and Analysis Branch (EV41) under the Contract MSFC (80NSSC18P1623).

\bibliographystyle{unsrt}
\bibliography{references}

\vspace{1cm}
\includegraphics[width=0.99653in,height=0.96944in]{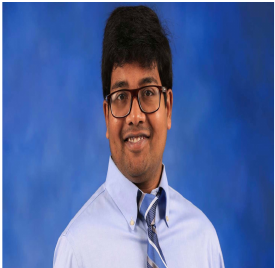}

\textbf{Sai Susheel Praneeth Kode} currently pursuing his Ph.D. in Electrical and Computer Engineering with concentration in aerospace dynamics and control systems from The University of Alabama in Huntsville (UAH), Huntsville, Alabama, USA. He received the BTech degree in Mechanical engineering from Jawaharlal Nehru Technological University Hyderabad (JNTUH), India and MS degree in Mechanical and Aerospace Engineering with concentration in autonomous robotic wheeled vehicles from UAH, in 2012 and 2015 respectively. After graduation with MSE degree, he worked as a New Product Introduction Project Engineer at General Electric (GE) Appliances, a Haier company at Lafayette, GA, USA from August 2015 to January 2018. His current research interests include sliding mode control and observation with applications to aerospace vehicle control.

\vspace{0.5cm}
\includegraphics[width=0.99653in,height=0.96944in]{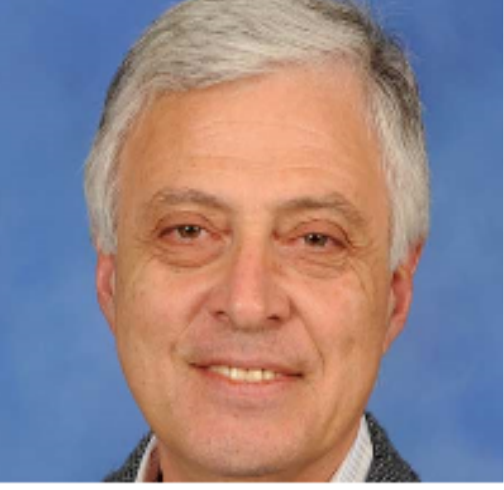}

\textbf{Yuri Shtessel} received the M.S. and Ph.D. degrees in Electrical Engineering with concentration in Automatic Control from the South Ural State University, Chelyabinsk, Russia in 1971 and 1978, respectively. Since 1993, he has been with the Electrical and Computer Engineering Department, The University of Alabama in Huntsville, and his present position is Distinguished Professor. His research interests include sliding mode control and observation with applications to aerospace vehicle and electric power system control. He is the author of more than 110 journal papers, 30 book chapters, over 230 papers in refereed conference proceedings, and 2 patents. Dr. Shtessel authored (with C. Edwards, L. Fridman, and A. Levant) a textbook ``Sliding Mode Control and Observation,'' Birkhauser, 2014. He is a recipient of the Distinguished Visiting Fellowship of the Royal Academy of Engineering, UK (2008); the Lady Davis Fellowship (2003), and the IEEE Third Millennium Medal (2000). He is a member of the \emph{IEEE Variable Structure Systems Technical Committee} and the \emph{IEEE CSS Conference Editorial Board}. Also, he serves as Subject Editor of the \emph{Journal of the Franklin Institute}, Technical Editor of \emph{IEEE Transactions on Mechatronics}, and Associate Editor of \emph{IEEE Transactions on Aerospace and Electronic Systems}. He also holds the ranks of Associate Fellow of AIAA and Senior Member of IEEE.
\end{document}